# Optics and bremsstrahlung estimates for channeling radiation experiments at FAST

J. Hyun[1], P. Piot[2, 3] and T. Sen[2],

[1]*SOKENDAI, Tsukuba, Ibaraki 305-0801, Japan*

[2] *Fermi National Accelerator Laboratory, Batavia, Illinois 60510, USA*

[3] *Northern Illinois University, DeKalb, Illinois 60115, USA*

## *Abstract*

This paper presents X-ray spectra of channeling radiation (CR) expected at the FAST (Fermi Accelerator Science and Technology) facility in Fermilab. Our purpose is to produce high brightness quasi-monochromatic X-rays in an energy range from 40 keV to 110 keV. We will use a diamond crystal and low emittance electrons with an energy of around 43 MeV. The quality of emitted X-rays strongly depends on parameters of the electron beam at the crystal. We present simulations of the beam optics for high brightness and high yield operations with bunch charges of 1 pC, 20 pC, and 200 pC. We estimate the X-ray spectra including bremsstrahlung background for a charge of 20 pC. The electron beam distributions with and without channeling in the diamond crystal are calculated. We discuss an X-ray detector system to avoid pile-up effect during high charge operations.

## *1. Introduction*

Channeling radiation (CR) can be generated when charged particles such as electrons and positrons pass through a single crystal parallel to a crystal plane or axis [1]. The main advantage of CR is to produce quasi-monochromatic high energy X-rays using a low energy electron beam below 100 MeV [2, 3, 4]. For instance, CR can emit X-rays with an energy of 16 keV using a 14.6 MeV energy electron beam [5]. By comparison synchrotron radiation (SR), currently the main X-ray source, requires a few GeV electron beam to generate X-rays of 16 keV. Moreover, intensities of CR are higher than those of PXR and transition radiation which are also produced using a crystal and an electron beam. Therefore, CR has the potential to be used for a compact X-ray source which can be disseminated to hospitals and laboratories.

Our purpose is to produce high brightness X-rays using a low-emittance electron beam, and to demonstrate that CR can be used as a compact high-brightness X-rays source. We plan to conduct CR generation experiments at the FAST (Fermi Accelerator Science and Technology) facility in Fermilab. The FAST injector which consists of a CsTe photocathode located in a 1+1/2-cell RF gun followed by two L-band (1.3 GHz) superconducting accelerating structures can generate a low emittance electron beam [6, 7]. The electrons energy can reach up to ~50 MeV downstream of the last superconducting cavity. A diamond single crystal with a thickness of 168 μm will be used because of its low atomic number Z, high Debye temperature, and large thermal conductivity. The crystal is oriented so that the electron beam propagates parallel to the (110) plane of the crystal. The expected spectra of CR at FAST were reported earlier [6, 8]. In this report, we present detailed calculations of the beam optics, background to CR from bremsstrahlung, electron beam distributions after the diamond crystal with and without channeling, and the X-ray detector system for the CR experiments.

In Section 2 of this paper, the energy, yield, and brilliance of CR are described. In Section 3, the FAST photoinjector is shown, where the details of the superconducting cavities, magnets and emittances for charges of 1 pC, 20 pC, and 200 pC are described. The beam optics and the X-ray spectra including bremsstrahlung background for each charge are shown in Section 4 and 5. The electron distributions after going through the crystal are shown in Section 6, and an X-ray detector system using Compton scattering to avoid the pile-up for high charge operations is described in the last section.

## *2. Channeling Radiation*

This section briefly presents the energy, yield, and brilliance of CR. When an electron beam travels with a sufficiently small transverse momentum with respect to a crystal plane, the electrons can be captured in bound states of the crystal's transverse potential, and consequently emit CR. Electron motion in the crystal is similar to that in an undulator, and the vibration period is very short, therefore, high energy X-rays can be emitted using a comparatively low energy electron beam. An important requirement for CR is that the electron beam divergence at the crystal must be smaller than the critical angle $\theta_c$ [10]. If beam divergences are larger than the critical angle, electrons in the crystal have too large a transverse momentum to be captured by the crystal potential; this is called de-channeling. The critical angle $\theta_c$ for CR is given by

$$\theta_c = \sqrt{\frac{2\gamma V_{max}}{mc^2(\gamma^2-1)}} \approx \sqrt{\frac{2V_{max}}{mc^2\gamma}}, \qquad (2-1)$$

where $V_{max}$ is the transverse potential of a crystal, $m$ is the electron mass, $c$ is the speed of light, and $\gamma$ is the Lorentz factor. For an electron energy of 43 MeV and a diamond crystal with the (110) plane, the critical angle $\theta_c = 1.1\ \text{mrad}$ using Eq. (2-1).

The energy spectrum of CR for electron energies below 100 MeV, derived by solving Schrodinger equation, is discrete [2, 3, 4, 5]. Electrons bound by the potential of a crystal plane or axis have discrete energy levels, and spontaneous transitions between the energy states generates quasi-monochromatic CR. On the other hand for electron energies above ~100 MeV, the energy spectrum of CR can be described by classical electrodynamics and it is broad and continuous [10]. The CR energies for electron energies below 100 MeV are given by [10]

$$E_\gamma = 2\gamma^2 \frac{(\varepsilon_i - \varepsilon_f)}{(1+\gamma^2\theta^2)}, \qquad (2-2)$$

where $\theta$ is the angle of the emitted photon from the incident electron, $\varepsilon_i$ and $\varepsilon_f$ are the energy eigenvalues of electrons in energy levels $i$ and $f$. This equation also shows that the CR energy can be tuned by changing the electron energy since $E_\gamma \propto \gamma^2$. The CR spectrum calculations reported here were done using a Mathematica notebook which was significantly corrected and modified from the published version [11], this modified version was used to successfully compare simulations with experimental results from a previous channeling experiment [8]. The CR energies for a diamond crystal and 43 MeV electron energy are shown in Table 2-1 for photons emitted at an angle of 0 degrees with respect to the incident electron beam.

Table 2-1: X-ray energies generated for a diamond crystal and 43 MeV electrons.

| Transition | X-ray energy [KeV] |
|---|---|
| $1 \rightarrow 0$ | 110 |
| $2 \rightarrow 1$ | 67.5 |
| $3 \rightarrow 2$ | 51 |

The photon yield for states $i \rightarrow f$ per steradian per photon energy per electron is written as [3, 8, 11,12]

$$\frac{d^2N}{d\Omega dE_\gamma}(i \rightarrow f) = \frac{\alpha_f \lambda_f^2}{\pi^{5/2}\hbar c} 2\gamma^2(\varepsilon_n - \varepsilon_m)\left|< \psi_f \left|\frac{\partial}{\partial x}\right| \psi_i >\right|^2 \times \int_0^d exp\left[-\mu\left(E_\gamma\right)(d-z)\right]P_n(z)$$

$$\times \int_0^\infty dt \frac{t^{-\frac{1}{2}}(1+2\alpha^2 t)(\Gamma_T/2)e^{-t}}{\left[(1+2\alpha^2 t)E_\gamma - E_0\right]^2 + [(1+2\alpha^2 t)(\Gamma_T/2)]^2}, \quad (2-3)$$

where $\alpha_f$ is the fine structure constant, $\lambda_c$ the Compton wavelength of the electron, $\varepsilon_n, \varepsilon_m$ the energies of the state $n, m$, $d$ the crystal thickness, $\mu\left(E_j\right)$ the photon absorption coefficient, $E_\gamma$ the photon energy, $\Gamma$ the total line width of the transition $n \rightarrow m$, $E_0$ photon energy at zero angle, $\alpha = \gamma\theta_{MS}$ $\theta_{MS}$: the multiple scattering angle, $P_n(z)$ the occupation probability of channeling state $n$ at crystal depth $z$, and $\psi_{m,n}$ the eigen function of state $m, n$. Also, $P_n(\theta)$ is [11, 13]

$$P_n(\theta) = |\langle\psi|exp(ik_x x)\rangle|^2 = \frac{1}{d_p}\left|\int_{-\frac{d_p}{2}}^{\frac{d_p}{2}} \psi_i exp(ik_x x)dx\right|^2. \quad (2-4)$$

One of the requirements for high intensity photons is to make $P_n(\theta)$ large. We calculated the occupation of different channeling states with Eq. (2-4). Figure 2-1 shows the initial populations as a function of the beam divergence at the crystal entrance for states of $n = 0, 1, 2,$ and $3$ when the electron beam is incident at a zero angle to the crystal. Black, red, blue, and green lines are for $n = 0, 1, 2,$ and 3, respectively. The beam divergence except for the state of $n = 1$ should be low to generate high intensity photons. In the state of $n = 1$, the initial population has a distribution with a peak at a beam divergence of about 0.35 mrad.

The photon yields for beam divergences of 0.1 mrad, 0.5 mrad, and 1 mrad at the crystal were calculated, and the results are shown in Figure 2-2. The energy widths are about 10 % for the different divergences, and photon intensities are higher for smaller beam divergences.

In general, the quality of X-ray sources such synchrotron radiation and XFEL is evaluated with the spectral brilliance [photons/s/mrad$^2$/mm$^2$/0.1% bandwidth]. The average brilliance of CR emitted from electrons can be expressed as [8]

$$B_{av} = \frac{I_{av}}{e} \frac{\gamma^2 Y (\sigma_e')^2 10^{-3}}{\varepsilon_N^2 \, \Delta E_\gamma / E_\gamma} Erf\left[\frac{\theta_c}{\sqrt{2}\sigma_e'}\right] \; photons/s - (mm - mrad)^2 - 0.1\%BW, \quad (2-5)$$

where $I_{av}$ is the average electron beam current, $e$ is the elementary electron charge, $Y$ is the total photon yield per electron, $\varepsilon_N$ is the normalized emittance, $\theta_c$ is the critical angle, see Eq. (2-1), $\Delta E_\gamma / E_\gamma$ is the relative width of the X-ray line, $\sigma_e'$ is the electron beam divergence, and $Erf$ is the error function. According to the Eq. (2-5), the average brightness is proportional to $1/\sigma^2$, which shows that beam sizes at a crystal location should be small to generate high brightness CR.

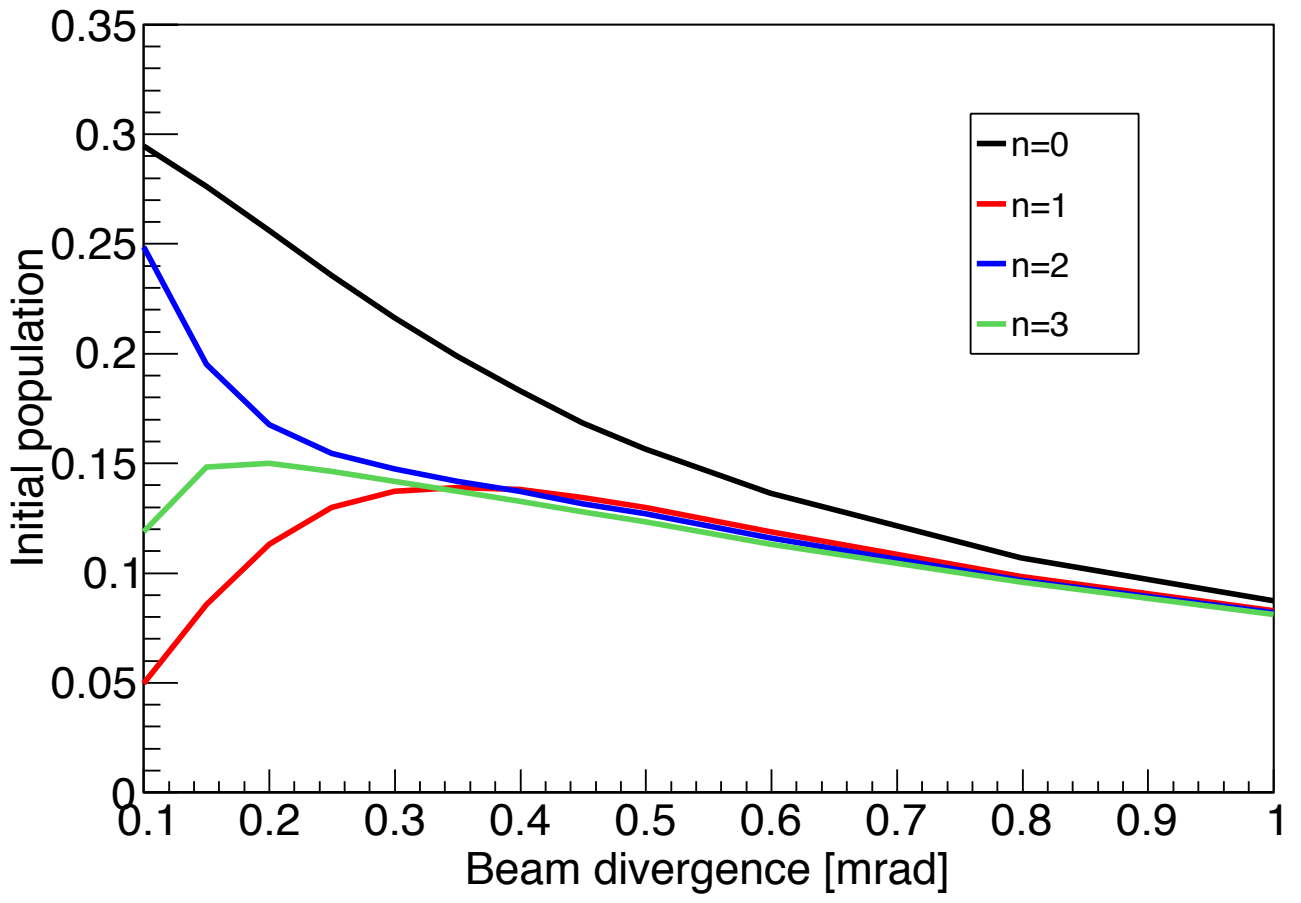

Figure 2-1: Initial population depending on beam divergences for different states of *n=0, 1, 2,* and *3* at incident angle of 0 degree.

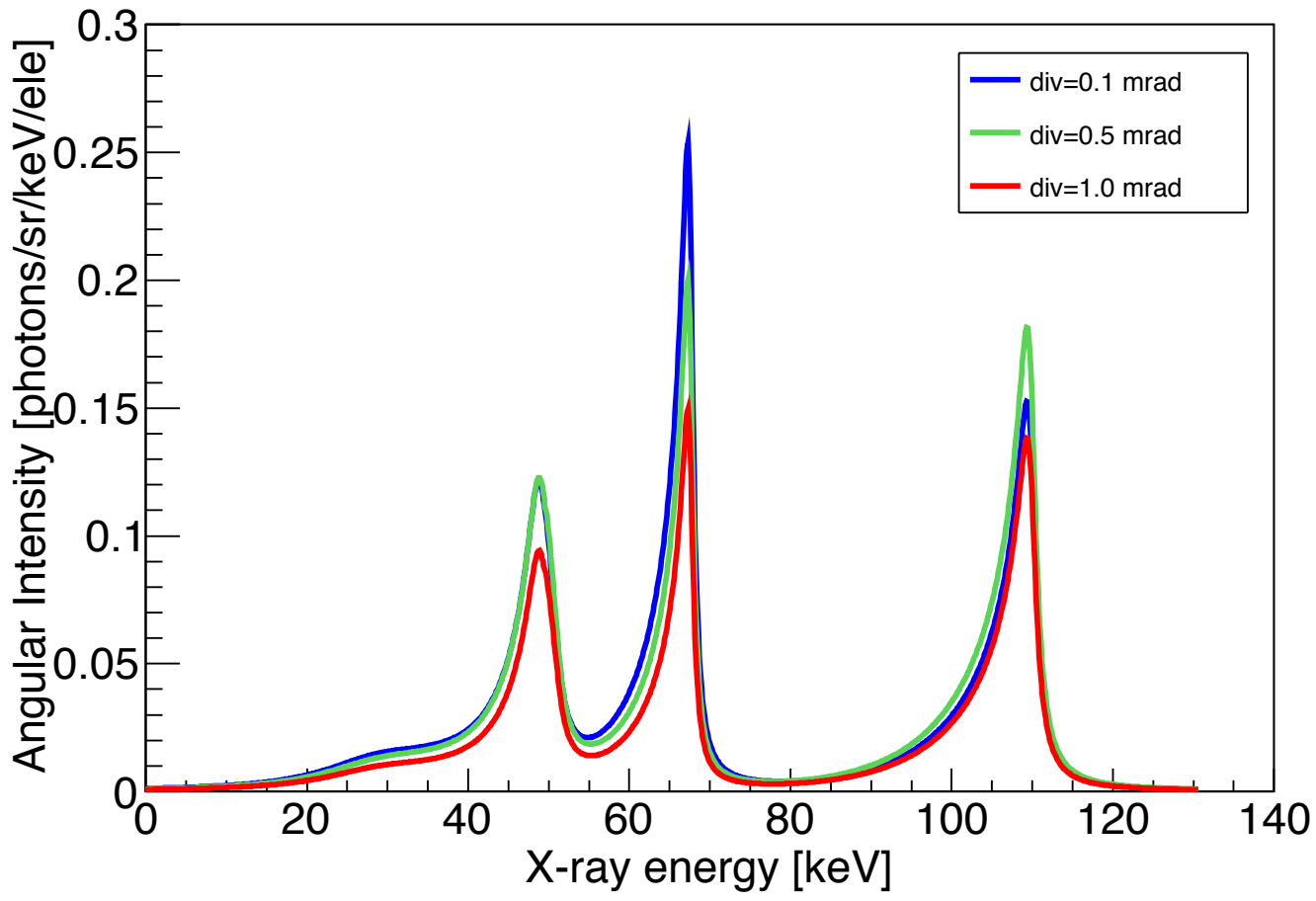

Figure 2-2: Photon yields for beam divergences of 0.1, 0.5, and 1.0 mrad.

## *3. FAST Photoinjector*

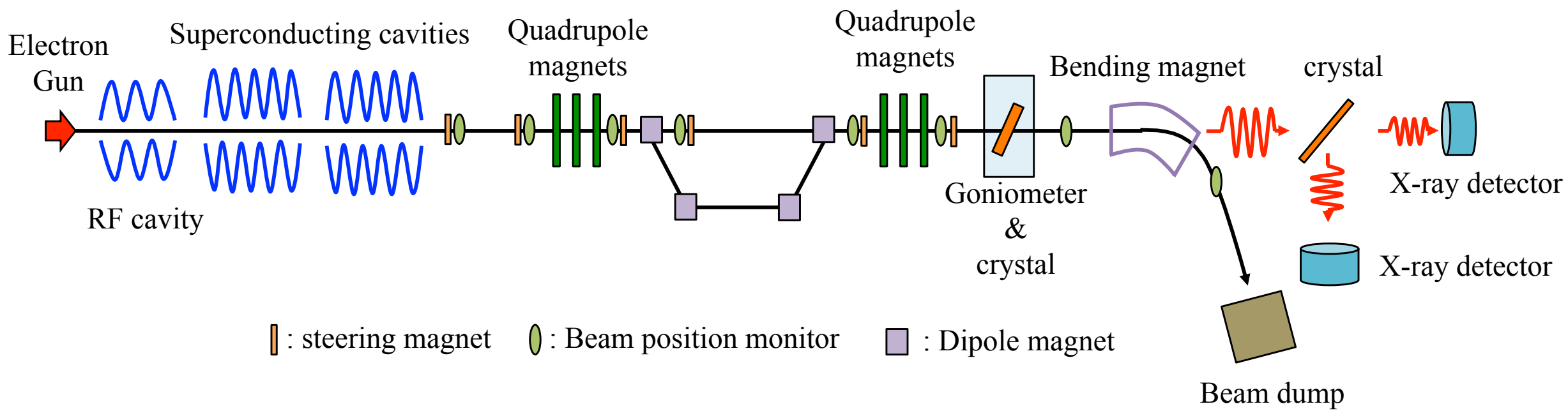

Figure 3-1: Layout of the FAST photoinjector

In this section, the FAST photoinjector and parameters of the electron beam for CR experiments are described. The main components in the beamline of this injector are a $Cs_2Te$ photocathode RF gun, two superconducting accelerating structures with TESLA style 9-cell cavities, quadrupole magnets, a chicane, a vertical bending magnet, and a beam dump [6]. Figure 3-1 and Table 3-1 show the layout and basic parameters of the photoinjector. The RF gun consists of a cathode with a molybdenum disk coated with $Cs_2Te$ mounted on the back plate of a 1+1/2-cell normal-conducting cavity operating at 1.3 GHz. The RF gun is identical to the one developed for the FLASH facility at DESY [7]. A bunch train repeated at 3 MHz with 1-ms duration is produced by irradiating the cathode with an ultraviolet laser pulse (wavelength of 263 nm). The electrons have an energy of ~5 MeV at the exit of the RF gun, and are accelerated up to energies in the range 43-50 MeV in the two superconducting structures operated at an RF frequency of 1.3 GHz.

In order to control the emittance growth at the source, two identical solenoid coils are utilized, a bucking coil surrounds the photocathode and the main solenoid with opposite polarity to the bucking coil is downstream of the bucking coil. The peak field is 0.28 T at the maximum current of 500 A. The bucking coil is used to cancel a magnetic fringe field from the main solenoid coil on the photocathode. The optimum coil currents are found by minimizing the measured beam emittance.

The goniometer housing the crystal and the X-ray detector for the CR experiment are located 17 m and 18.5 m, respectively from the photocathode. The goniometer stage can be rotated around vertical and horizontal axes for an incident electron beam direction and can slide horizontally. The goniometer is equipped with a target holder which houses a clear aperture, the diamond crystal and a 50 micron thick Al foil. The foil can be used as a beam monitor and for beam alignment. The hole is used when the crystal is not needed.

The electrons after going through the crystal are bent by the dipole magnet, positioned downstream of the goniometer, into the beam dump. This dipole magnet provides a kick of 22.5 degrees for transporting the electron beam into the beam dump. At 50 MeV beam energy, this requires an integrated field of 640 G-m. The quadrupole magnets used to control electron beam sizes in the beamline have a maximum gradient of 6.6 T/m at an energy of 50 MeV, their bore diameter is 54.6 mm and the effective magnet length is 617 mm. Also, eight steering magnets are inserted in the beamline to correct the electron beam trajectory. Each steering magnet is capable of a maximum kick of 7.5 mrad to a 50 MeV electron beam. Table 3-2 shows magnet parameters for a 50 MeV electron beam.

The chicane displayed in Figure 3-1 is commonly used for bunch compression and energy collimation. In the FAST photoinjector, the four dipole magnets respectively provide a bending angle of (+, -, -, +) 18 degrees yielding a longitudinal dispersion of $R_{56} = -0.18\ m$. In the CR experiment, the main purpose of using the chicane is to collimate away dark current of lower energy than the main photocathode current.

Several yttrium aluminum garnet (YAG) screens and electromagnetic beam-position monitors (BPMs) are available along the beamline to respectively measure the beam transverse profile and beam centroid position. Two YAG screen monitors are positioned before and after the goniometer, and we estimate the beam sizes at the crystal by using these two monitors and the transfer matrix. Also, a screen monitor inserted downstream of the vertical-bending magnet is used to obtain the beam energy and the energy spread by measuring the beam size and the vertical shift of the electron beam. In general, electron momentum $P$ is given by

$$P\left[\frac{GeV}{c}\right] = 0.3B\rho\ [Tesla - m]\ \ (3-1)$$

where $B$ is the magnetic field and $\rho$ is the bending radius. Electron beam energy can be computed from Eq. (3-1) with the $B$ ($\propto$ current) and the bend radius of the vertically bending dipole magnet when the electron beam is observed on the YAG screen monitor. The rms beam size $\sigma$ is determined from

$$\sigma = \sqrt{\varepsilon\beta + (\eta\ \delta_p)^2} \qquad (3\text{-}2),$$

where $\varepsilon$ is the beam emittance, $\beta$ is the betatron function, $\delta_p$ is the rms fractional momentum spread, and $\eta$ is the dispersion which can be computed. When the beta function is small at the YAG screen, the rms beam size is

$$\sigma \approx |\eta\delta_p|\ \ (3-3)$$

Therefore, the energy spread can be measured by minimizing the beam size (beta function) at the YAG screen downstream of the dipole magnet.

Electron beam dynamics from the photocathode to downstream of the second superconducting cavity (8m) were simulated, including space charge effects, using the tracking program ASTRA [15] for bunch charges ranging from 1 pC - 3.2 nC. Table 3-3 shows the Twiss parameters, the normalized emittances, and the energy spreads for 1, 20, and 200 pC.

Table 3-1: Basic parameters of the FAST's photoinjector.

| Parameter | Value |
|---|---|
| Beam energy | 43 -50 MeV |
| Bunch charge | 1-200 pC |
| Bunch frequency | 3 MHz |
| Normalized emittance | 0.52 mm-rad |
| Bunch length | 3 pC |
| Energy spread | < 0.2% |

Table 3-2: Magnet parameters for 50 MeV electron beam.

| Magnet | Max. current | Max Magnet field | Kick angle | Length |
|---|---|---|---|---|
| Solenoid coil (main and bucking) | 500 A | 0.28 T | ---- | ---- |
| Steering magnet (for orbital correction) | 7.6 A | 12.5 G-m Integrated field | 7.5 mrad Max. at 50 MeV | ---- |
| Bending magnet (for beam dump) | 7.7 A | 640 G-m Integrated field | 22.5 degrees | 265.3 mm (effective) |
| Dipole magnet (for chicane) | ---- | ---- | 18.0 degrees | 264.7 mm (effective) |
| Quadrupole magnet | 9.0 A | 6.6 T/m 1.1 T at pole tip | ---- | 167 mm, 54.6 mm (effective, bore) |

Table 3-3: Twiss parameters, normalized emittances, and energy spread for 1, 20, and 200 pC at 8 m from the photocathode, from ASTRA simulations.

| Charge [pC] | ε [μm-rad] | αx (=αy) | βx (=βy) [m] | ΔE/E [%] |
|---|---|---|---|---|
| 1 | 0.02 | -43.3 | 309.5 | 0.1 |
| 20 | 0.19 | -3.8 | 21.3 | 0.1 |
| 200 | 0.52 | -3.6 | 18.9 | 0.2 |

## 4-1. Beam optics solutions for CR

In this section, we present optics solutions for charges of 1, 20, and 200 pC (1) high brightness solutions with low beam sizes and divergences close to the critical angle (consistent with the beam emittance) and (2) high yield solutions with a large beam size and low divergences at the crystal.

### *4-1-1. Beam optics solutions for high brightness CR*

(1) For bunch charges of 1, 20, and 200 pC, the beam optics from downstream of CC2, the second superconducting cavity, to the beam dump was simulated with initial parameters in Table 3-3 using SAD (Strategic Accelerator Design) computer code [14]. Figure 4-1 shows the optics functions obtained by minimizing the beam size at the crystal for bunch charges of 1, 20, and 200 pC. Green and red boxes show quadrupole magnets and bending magnet respectively, and blue and red lines represent the horizontal and vertical planes respectively. The optics matching was done with two sets of triplet magnets; (Q108, Q109, Q110) and (Q118, Q119, Q120). Their polarities are (+, -, +) and (+, -, +); + and - means horizontal and vertical focusing, respectively. The quadrupole magnets used in matching are constrained to a maximum gradient of 6 T/m. Starting with the initial conditions shown in Table 3-3, the minimum beta functions at the crystal are 0.3 mm, 0.3 mm, and 0.5 mm
for 1, 20 and 200 pC. Table 4-1 shows the minimum beam sizes, beam divergences and beta functions at the crystal for three charges. The beta functions (and beam sizes) are largest in the final quadrupole magnets before the crystal, however, the beam sizes are much smaller than the radius of the beam pipe.

The calculated beam sizes ($\sigma_{x,y} = \sqrt{\varepsilon_{x,y}\beta_{x,y}}$) for minimum beta functions are quite different from those obtained by particle tracking. The racking includes chromatic effects. Figure 4-2 shows beam sizes at the crystal as functions of the beta functions for the different charges calculated with both methods. The dots are the beam sizes obtained by particle tracking, and lines are the beam sizes from the analytical calculation. When an electron beam is strongly focused by quadrupole magnets, the focus point is different for different electron energies because a low energy electron is bent more than a high energy electron. As the result of these chromatic aberrations, the transverse beam sizes at the crystal become larger than those obtained by analytical calculation. Assuming sextupole magnets cannot be placed in the chicane to correct the chromaticity, the only other option to keep chromatic effects small is that the electron beams have a small enough energy spread. Figure 4-3 shows the beam sizes at the crystal depending on momentum spread simulated by particle tracking when the beta functions for 20 pC are 3 mm at the crystal. The energy spread affects the horizontal size more due to strong horizontal focus in the last triplet magnet. Also, this shows that momentum spread should be less than 0.1% at the crystal. Expected energy spread for (1, 20) pC is about 0.1 % but larger for 200 pC. Therefore, the effect of the chromatic aberration should not be significant for the two lower chargers.

The beam divergences at the crystal for 1, 20 and, 200 pC are 0.3, 0.9, and 1.1 mrad, respectively. For the different charges, 94% (1 pC), 50% (20 pC), and 34 % (200 pC) of particles in a bunch satisfy the channeling condition of the critical angle, and the other particles pass through the crystal without channeling. Although the brilliances decrease slightly, the losses have little effect on the brilliances because it depends on the square of the beam size at the crystal.

Table 4-1: Minimum beam sizes, beam divergences, and beta functions at the crystal for the different charges. Initial conditions are shown in Table 3-3.

| Charge [pC] | $\beta_x$ (=$\beta_y$) [mm] | $\sigma_x$, $\sigma_y$ [μm] | $\sigma_x'$, $\sigma_y'$ [mrad] |
|---|---|---|---|
| 1 | 3 | (1.3, 1.3) | (0.3, 0.3) |
| 20 | 3 | (4.1, 4.0) | (0.9, 0.9) |
| 200 | 5 | (9.6, 9.6) | (1.1, 1.1) |

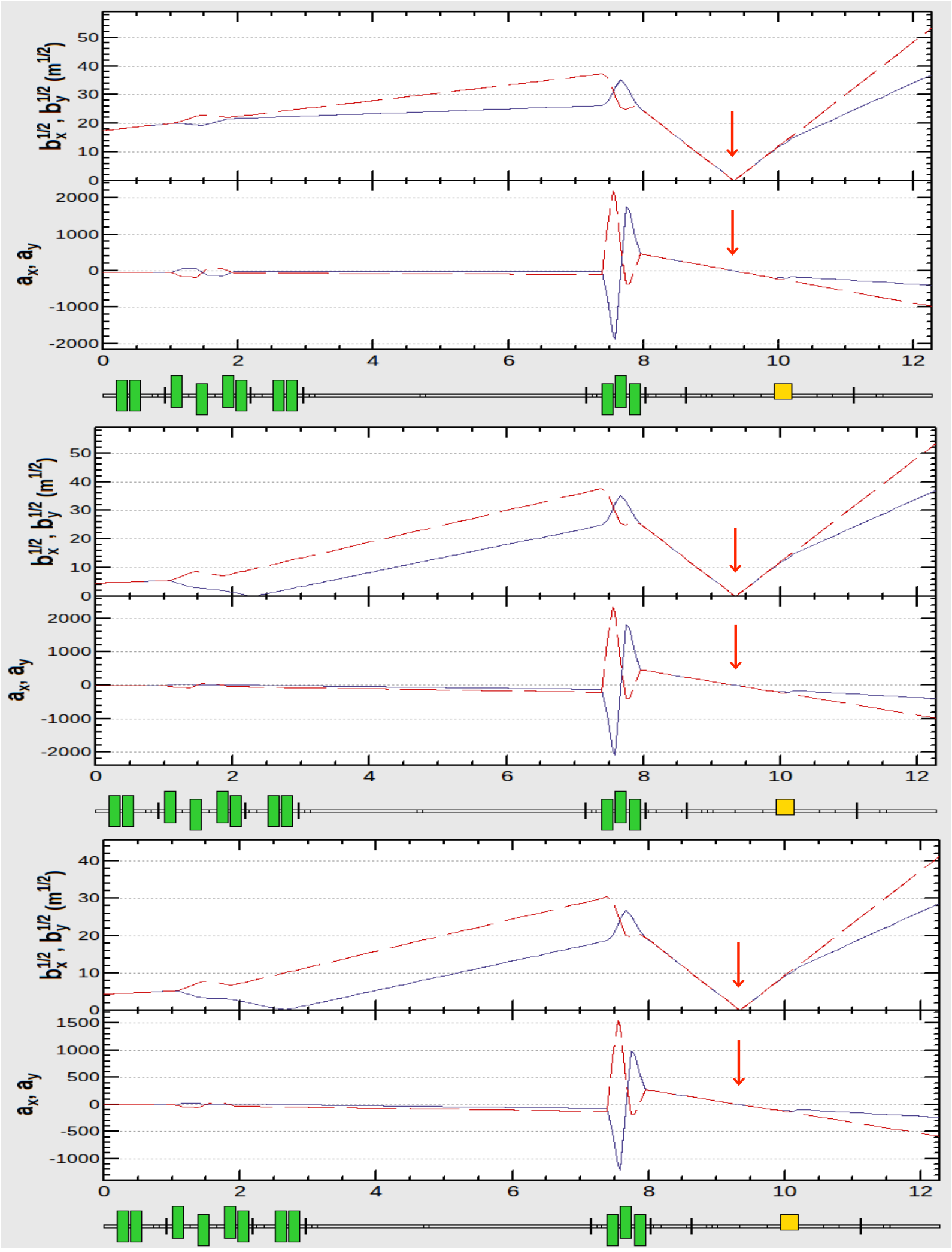

Figure 4-1: Optics functions of the transport line when beam sizes at crystal are minimum for 1 pC, 20 pC, and 200 pC. Top: 1 pC, Middle: 20 pC, Bottom: 200 pC. Blue and red lines show the horizontal and vertical planes. Green and yellow boxes represent quadrupole and dipole magnets respectively.

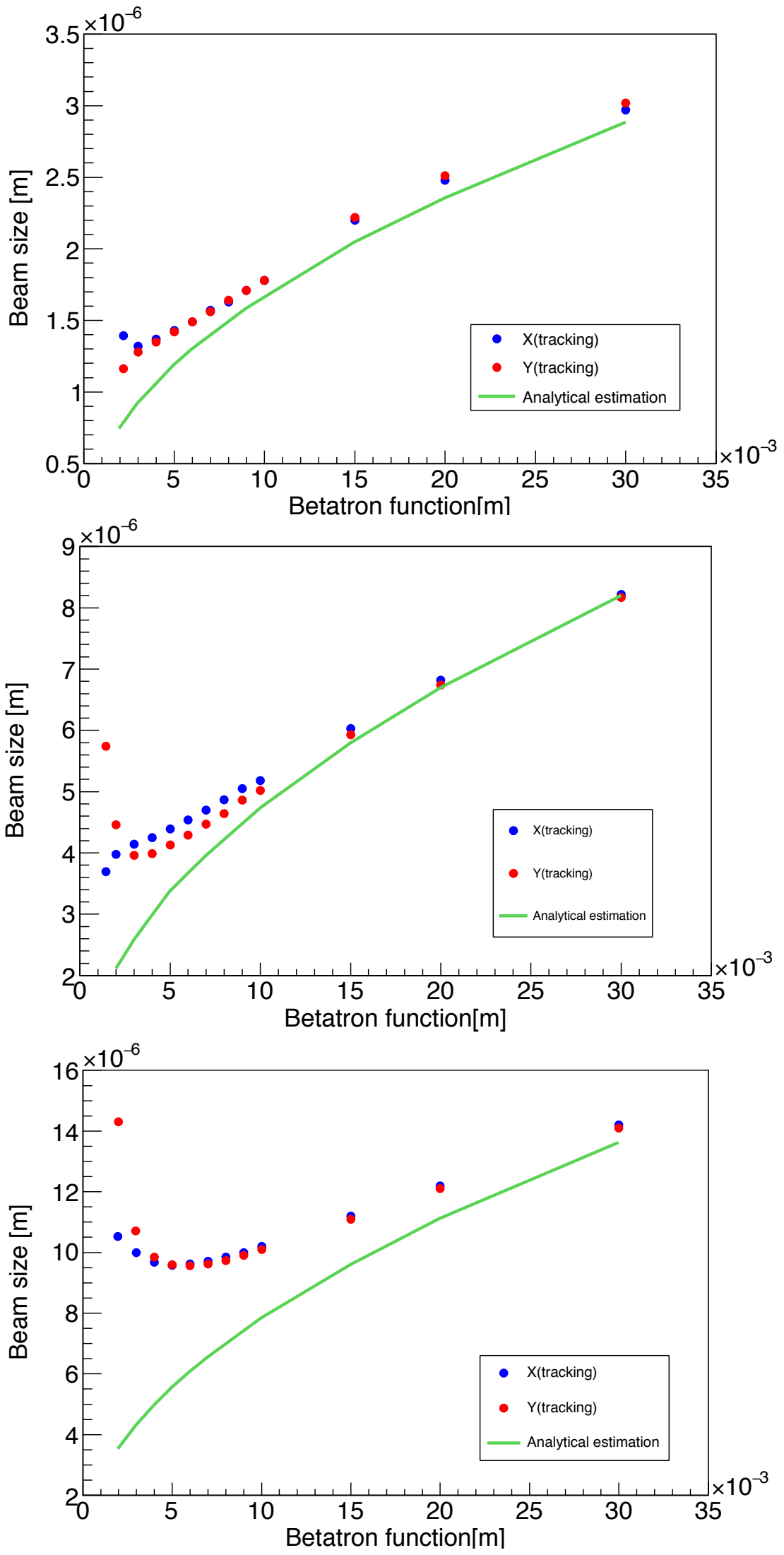

Figure 4-2: Beam sizes at the crystal depending on the beta functions for 1 pC, 20 pC, and 200 pC. Top: 1 pC, Middle: 20 pC, Bottom: 200 pC. Blue and red dots show horizontal and vertical beam sizes obtained from particle tracking. Green lines are the beam sizes from analytical calculation without chromatic effects. The rms energy spreads for these bunch charges are shown in Table 3-4.

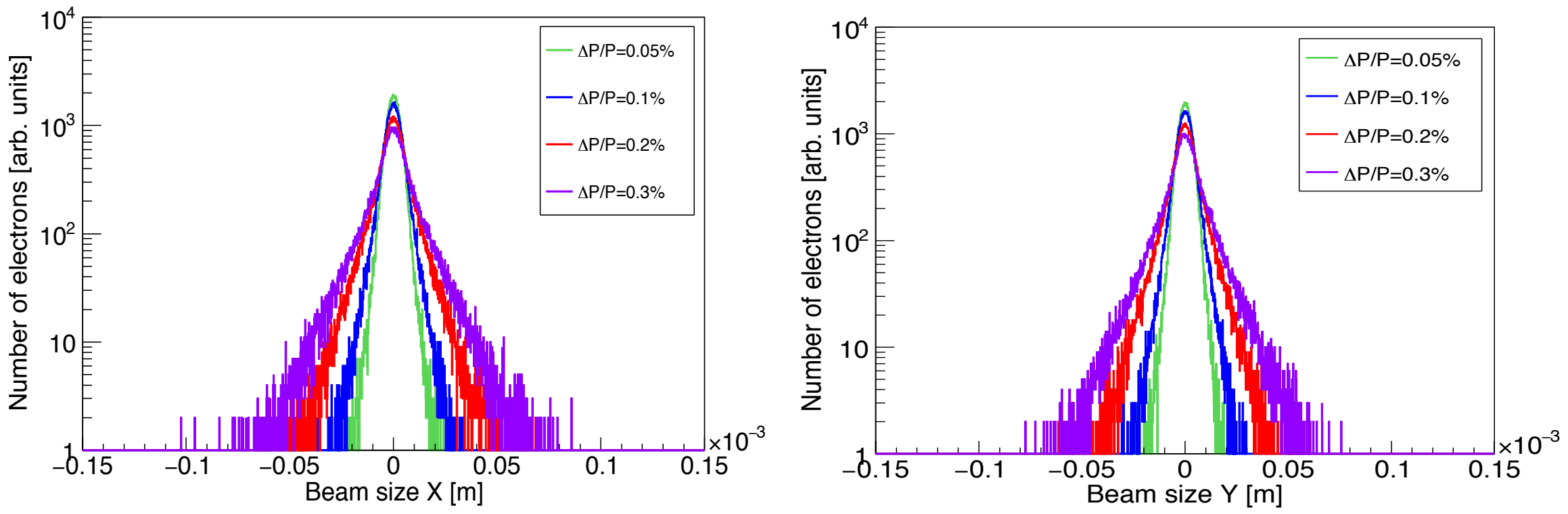

Figure 4-3: Beam size distributions at the crystal for different momentum spreads for 20 pC when the beta functions at the crystal are 0.3 mm. Left: horizontal plane, right: vertical plane.

### *4-1-2. Beam optics solutions for high yield CR*

(2) To produce high yield CR, the beam divergence at the crystal should be much smaller than the critical angle from Eq. (2-3) and (2-4) in Sec.2. This means that beam size at the crystal will be large due to the conservation of beam emittance. Hence the chromatic aberration can be ignored (see Figure 4-2). We simulated the beam optics so that the electron beam has divergences of 0.1 mrad and there is an optics waist (twiss parameter $\alpha_{x,y} = 0$) at the crystal for each charge. The optics functions along the beamline are shown in Figure 4-4. The beam optics simulated for each charge is different due to different initial twiss parameters but shows same behavior. For charges of 1 pC, 20 pC, and 200 pC, the beam sizes at the crystal are 2.9 μm, 22 μm, and 61 μm, respectively. Table 4-2 shows beam sizes and beta functions for a beam divergence of 0.1 mrad at the crystal. The beam sizes obtained from analytical calculation of the betatron beam size without chromatic effects are consistent with those from particle tracking.

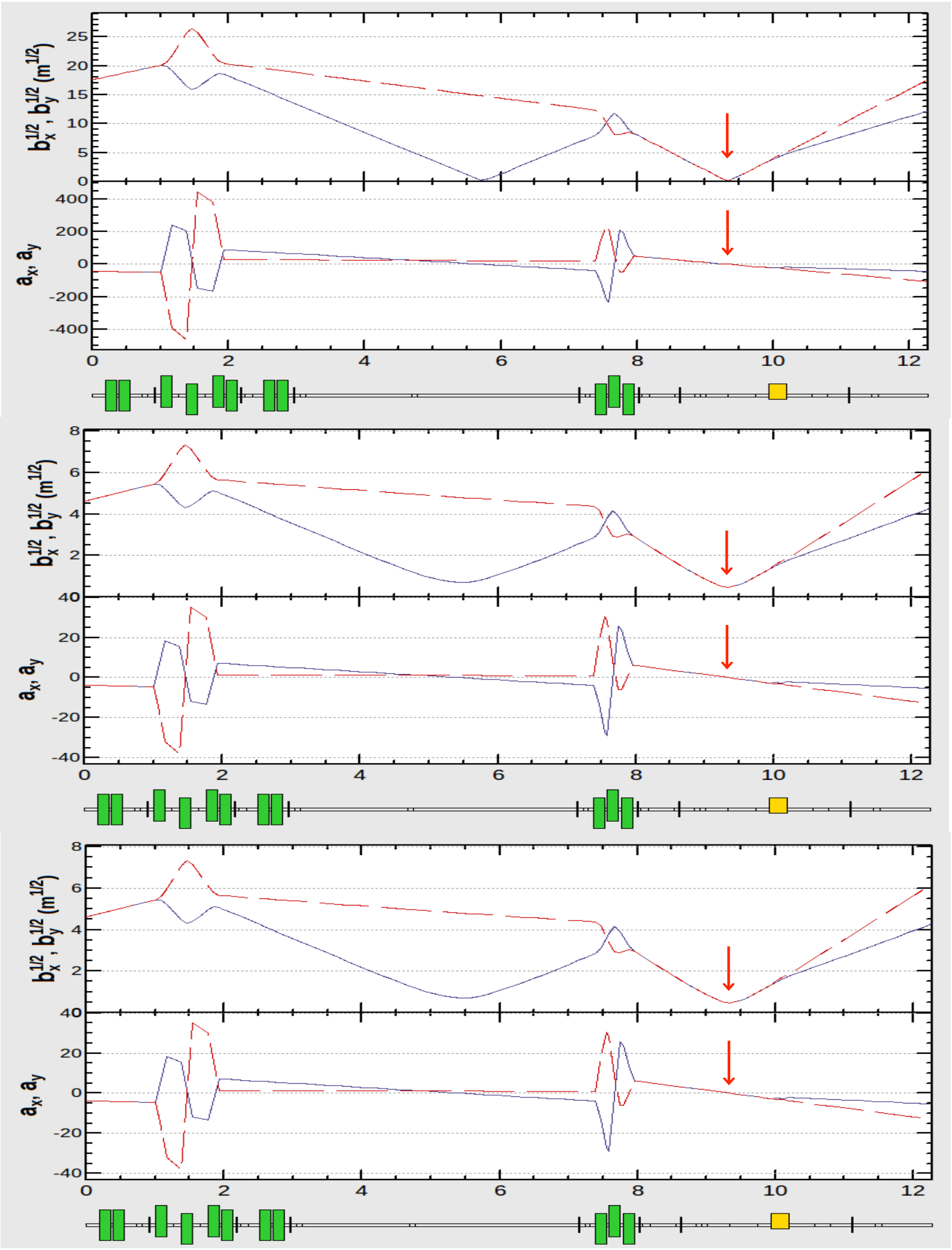

Figure 4-4: Beam optics for the high yield solution along the transport line from the last cavity to the beam dump. The beam divergences at the crystal are matched to 0.1 mrad in each case. Top: 1 pC, Middle: 20 pC and Bottom: 200 pC. Blue and red lines show the horizontal and vertical planes. Green and yellow boxes represent quadrupole and dipole magnets respectively.

Table 4-2: Beam sizes and beta functions for a beam divergence of 0.1 mrad at the crystal.

| Charge [pC] | $\beta_x$ (=$\beta_y$) [m] | $\sigma_x$=$\sigma_y$ [μm] |
|---|---|---|
| 1 | 0.03 | 29 |
| 20 | 0.22 | 22 |
| 200 | 0.61 | 61 |

## 4-2 Beam optics solutions with misalignment and magnetic field error

In this section, we discuss error corrections of quadrupole magnets with misalignments and magnetic field errors. These magnets with misalignments and magnetic errors, give the beam dipolar and quadrupolar kicks with different strengths, and cause different focusing, beam deflections and excitations of betatron oscillations and unwanted dispersions. Therefore, beam optics must be simulated including these effects. In the FAST linac, corrector magnets such as steering magnets and skew magnets for corrections of the errors have been installed.

Misalignments and magnetic errors, assuming that they follow a Gaussian random distribution with a cutoff of 3σ, are set in all quadrupole magnets, and the optics calculations are done with 10 different error seeds. The RMS values of misalignments and gradient errors set in the quadrupole magnets are shown in Table 4-3. We assume that the rms errors are as follows: alignment: 0.2 mm, gradient: 0.3%, and rotation: 0.2 mrad in each plane. The orbits are corrected with 7 steering magnets and 11 BPMs. Figure 4-5 shows the central orbits for the small beam size and the small divergence solutions with and without orbit corrections for a charge of 200 pC. Green lines show the central orbit without the correction. When the orbits are not corrected with the steering magnets, the horizontal orbit error is over 3 mm in the second triplet where beam sizes are maximum and the vertical orbit error is about 2 mm at the crystal.

Figure 4-6 shows the central orbits of electron beam for the small beam size and the small divergence solutions along the beamline after the orbit correction. Blue and red lines are the average orbits of horizontal and vertical, respectively. The error bars are the standard deviations of orbits. After the orbit correction, the deviations of the two solutions from a central orbit at the crystal can be reduced to about ±0.2 mm for 200 pC. Table 4-4 shows beam sizes and beta functions achieved at the crystal for the small beam solutions using quadrupole magnets with errors. The beam sizes at the crystal for 200 pC are about 11 μm.

Table 4-3: Alignment and field errors for the quadrupole magnets.

| Alignment error | | Gradient error | Rotation error |
|---|---|---|---|
| ΔX [mm] | ΔY [mm] | ΔK/K | Δθ [mrad] |
| 0.2 | 0.2 | 0.003 | 0.2 |

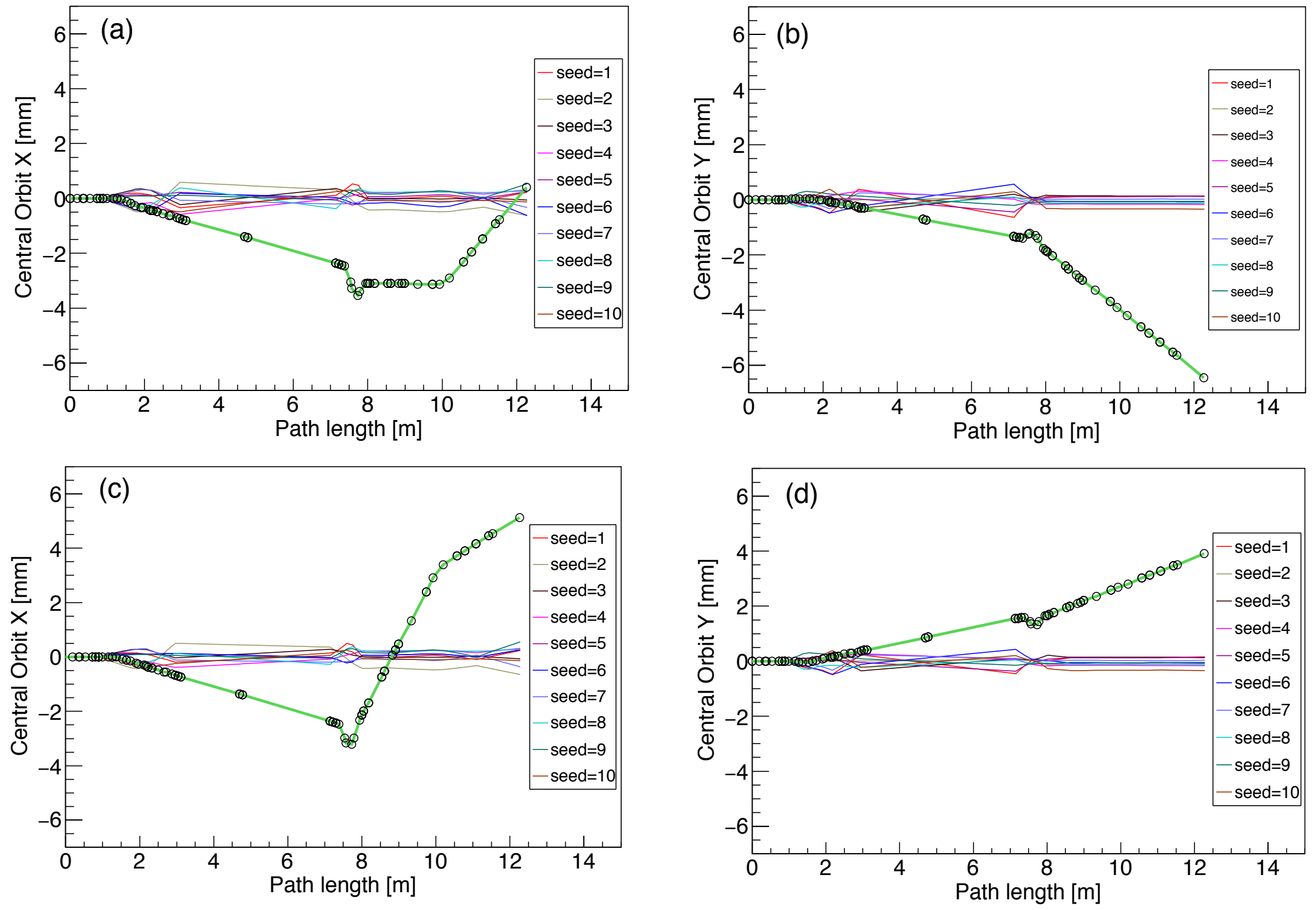

Figure 4-5: Orbit corrections with steering magnets for the small beam size (upper plots (a) and (b)) and the small divergence solutions (bottom plots (c) and (d)) for 200 pC. Left: horizontal, Right: vertical. Green lines show before corrections. 10 seeds are used

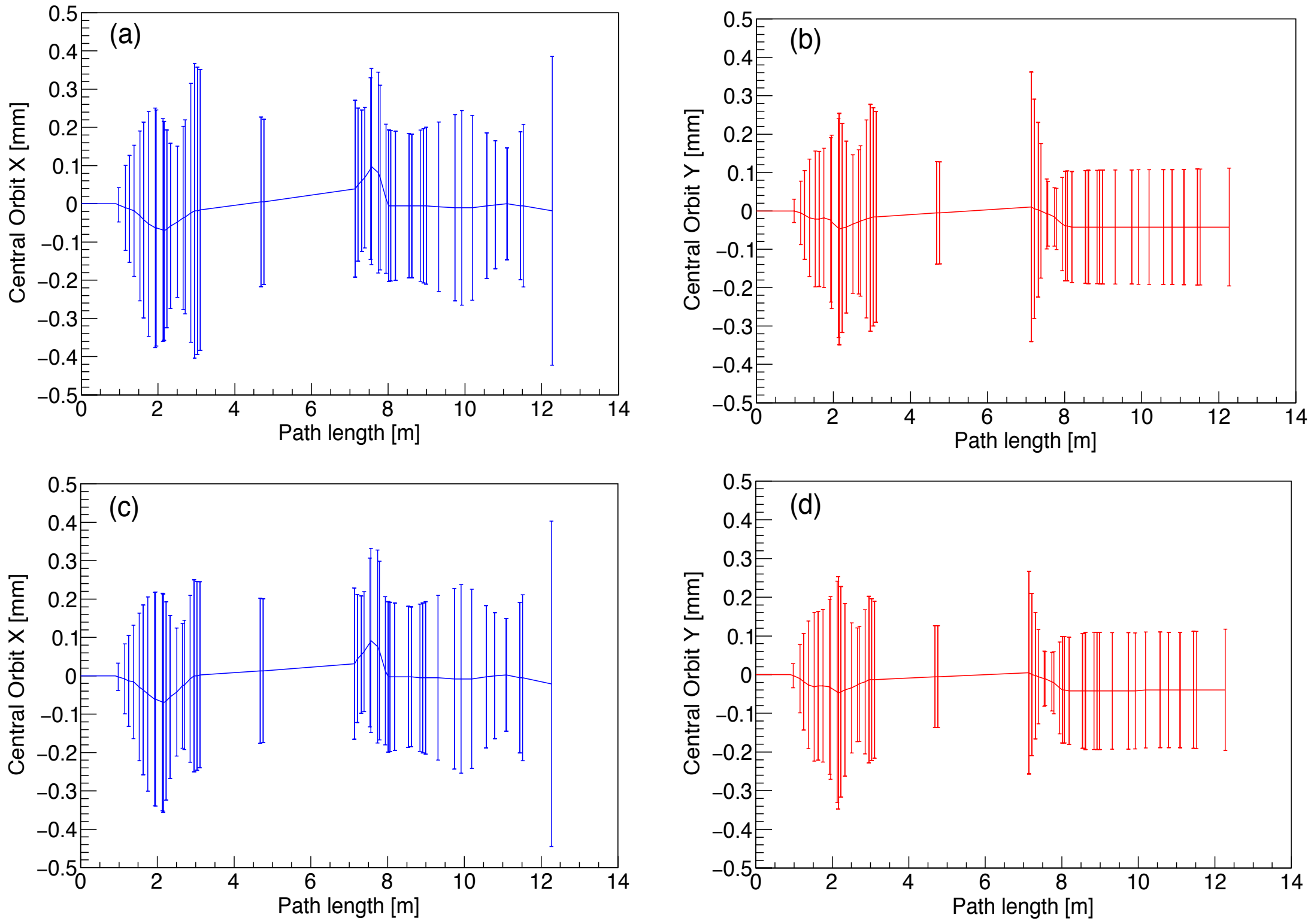

Figure 4-6: Transverse central orbit along the beamline for the small beam size (upper plots (a) and (b)) and the small beam divergence solutions (bottom plots (c) and (d)) after the orbit correction for 200 pC. 10 seeds are used. Blue and red lines show average orbits in the horizontal and vertical planes, and black lines show the standard deviations for 10 seeds.

Table 4-4: Beam sizes and beta functions at the crystal achieved for the small beam size solution using the quadrupole magnets with the misalignments and field errors.

| Charge [pC] | $\beta_x$ (=$\beta_y$) [mm] | $\sigma_x$, $\sigma_y$ [μm] |
|---|---|---|
| 1 | 5 | (1.8±0.4 , 1.7±0.3) |
| 20 | 6 | (5.2±0.8 , 5.0±0.8) |
| 200 | 8 | (10.5±0.9 , 10.3±0.8) |

## *5. Background from bremsstrahlung*

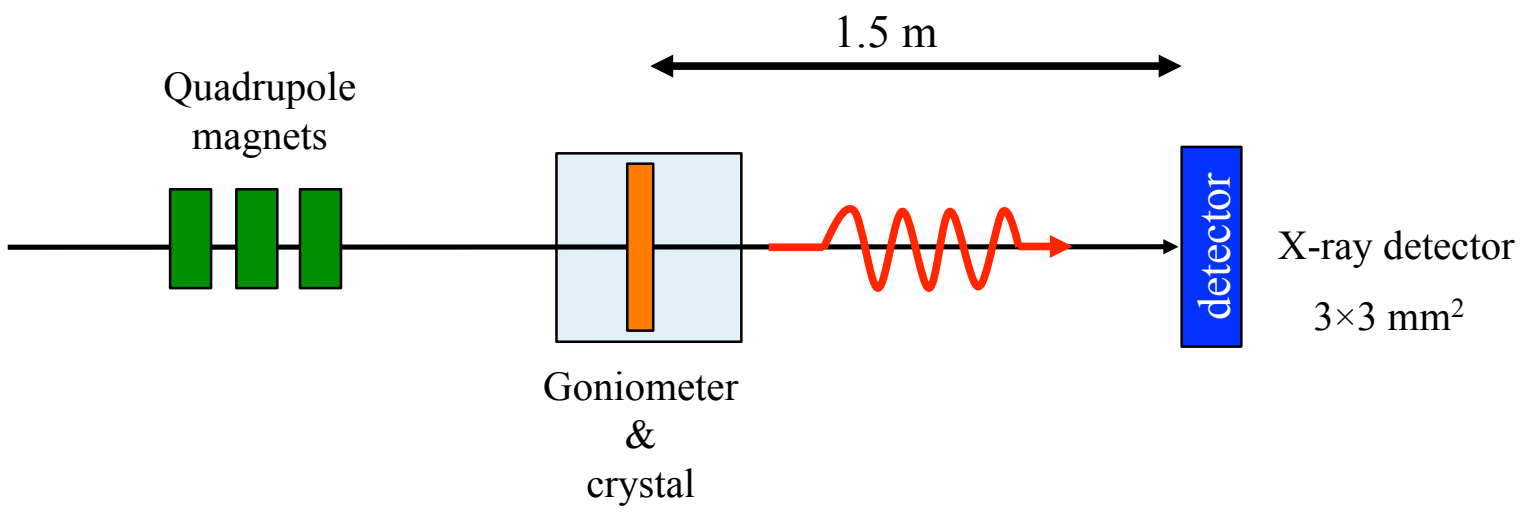

Figure 5-1: Layout showing goniometer and the detector.

Bremsstrahlung (BS) X-rays are produced when relativistic electrons passing through a crystal are scattered by the atomic nuclei. The spectrum of BS is continuous, covering a wide energy range from microwave to hard X-rays. The maximum BS energy can extend to nearly the electron energy. BS is generally background in CR experiments since it is emitted in the same forward direction as CR with respect to the incident electron beam. In order to examine the effect of BS in our CR experiment, we estimated the photon count and the energy distribution of the background registered in the detector with Geant4 [16] simulations. The X-ray detector with an aperture of 3x3 mm$^2$ is assumed to be located at 1.5 m from the target crystal so that the detector's acceptance is 2 mrad. The layout showing the X-ray detector and simulation geometry is shown in Figure 5-1. In the simulations, $10^8$ particles (20 pC/bunch) with an energy of 43 MeV and a diamond crystal with a thickness of 168 μm are used.

Figure 5-2 shows the scattering angles of photons emitted from electron beams going through the diamond crystal in the forward direction and in the detector. The left plot in Fig. 5-2 shows the angular distribution in the forward direction, while the right plot shows the angular distribution in the detector, and the red line shows the Gaussian fit. The standard deviation of the angular distribution of photons emitted in the forward direction is about 49 mrad, while the rms beam size obtained from the Gaussian fit is $1\sigma = 13$ mrad, which is good agreement with the BS scattering angle $1/\gamma = 12$ mrad. This shows that the angular distribution of all the scattered BS photons is very non-Gaussian, but the distribution within the detector's acceptance is nearly uniform.

From Fig. 5-2, about $10^6$ photons (or about $10^{-2}$ photons/electron) are emitted in the forward direction, and ~$10^4$ photons of them go into the detector. Also, the photon count hitting the detector as a function of beam size for different beam divergences at the crystal is plotted in Figure 5-3. For the electron beam size below 0.1 mm at the crystal, the number of photons going into the detector is substantially constant at ~$10^4$ particles. However, it decreases when the electron beam size is over 0.1 mm. The photon beam size $\sigma_2$ at the X-ray detector can be written as

$$\sigma_2 = \sqrt{\sigma_1^2 + L^2{\sigma_1}'^2}, \cdots (5-1)$$

where $\sigma_1$ is the photon beam size and ${\sigma_1}' (= 13\, mrad)$ is the photon beam divergence at the diamond crystal, and $L\ (= 1.5\, m)$ is the distance between the crystal and the X-ray detector. The number of photons per beam area at the X-ray detector location is proportional to $\frac{1}{\pi{\sigma_2}^2} (= 1/\pi(\sigma_1^2 + L^2{\sigma_1}'^2))$. Therefore, the photon count $I$ going into the X-ray detector with the detector area $S (= 9\, mm^2)$ is

$$I \propto \frac{S}{\pi{\sigma_2}^2} = \frac{S}{\pi(\sigma_1^2 + L^2{\sigma_1}'^2)}. \cdots (5-2)$$

This photo count $I$ is shown as a function of the beam size at the crystal in the right plot in Figure 5-4. The number of photons hitting the detector decreases for the beam size $\sigma_1$ of over 0.1 mm, which is consistent with simulation results with Geant4 shown in the left plot in Figure 5-4. In our CR experiment for an electron bunch charge of 20 pC, we expect about $10^4$ background BS photons per electron bunch will be registered in the detector, since the beam size at the crystal is desired to be below 0.1 mm. Also, the ratio of background photons in the detector to incident electrons is approximately $10^{-4}$, therefore, about 5x$10^2$ and $10^5$ photons per bunch for bunch charges of 1 pC and 200 pC respectively will be registered as background. Table 5-1 shows the expected number of background photons registered in the detector.

Table 5-1: The number of background photons registered in the detector for three charges.

| Charge [pC] | Number of photons |
|---|---|
| 1 | ~5x$10^2$ |
| 20 | ~$10^4$ |
| 200 | ~$10^5$ |

The BS energy distributions in the forward direction and in the detector are shown in Figure 5-4. The energy distributions for the two cases have the almost same behavior, which indicates that the BS photon energy does not depend on its emitted angle. Moreover, in order to investigate the dependence of the BS energy distributions on electron energies, we simulated them with electron energies of 43 MeV, 60 MeV, 80 MeV, and 100 MeV using Geant4. The number of photons and the energy spectra in the forward direction are shown in Table 5-2 and Figure 5-5, respectively. The photon distributions are shown to a maximum of 200 keV (left plot) and 100 MeV (right plot). The left plot shows that maximum yield occurs close to 10 keV for all the electron energies and the distribution from about 40 keV to 200 keV is nearly the same for the different electron energies. The right plot shows that the photon spectra extend to the electron beam energy in all cases. Table 5-2 shows that the total number of BS photons produced in the forward direction is unchanged for different electron energies, because the differential cross section $\chi$ of BS can be approximately written as [17]

$$\frac{\partial\chi}{\partial(\hbar\omega)} \cong \frac{16}{3}\alpha r_e Z^2 \ln\left(\frac{233}{Z^{1/3}}\right), \cdots (5-3)$$

where $\alpha$ is the fine-structure, $r_e$ is the classical electron radius, and $Z$ is the atomic number of the crystal. This shows that the photon yield, obtained by integrating over all photon energies, is independent of the electron energy.

The expected channeling spectra, including the background, when a 43 MeV electron beam is incident on a 168 μm thick diamond crystal parallel to the (110) planes, are shown in Figure 5-6. The CR yields without the background are calculated with Eq. (2-3), and then the process of de-channeling and re-channeling in a crystal are taken into account. This model affects populations in bound states and leads to reduce photon yields. The CR photon count including this process can be calculated using a free parameter $n_f$. The appropriate $n_f$ was decided from photon yields obtained in CR experiments at the ELBE facility [8]. The "CR_high" label in Fig. 5-6 corresponds to the result for a case without dechanneling $n_f = 21$, "CR_mid" is the result for the case $n_f = 17$ estimated from the experimental values, and "CR_low" is a more conservative result with $n_f = 13$ corresponding to a lower yield. The X-rays with discrete energies of 110 keV (transition: 1→0), 67.5 keV (transition: 2→1) and 51 keV (transition: 3→2) are generated at the angle of 0 degree. The ratios of CR signal for CR_mid $(n_f = 17)$ to the BS background are about 7 at 110 keV, 5 at 67.5 keV, and 2 at 51 keV. These theoretical values of signal to backgrounds imply that the CR signal should be clearly observable at the higher energy CR spectral lines.

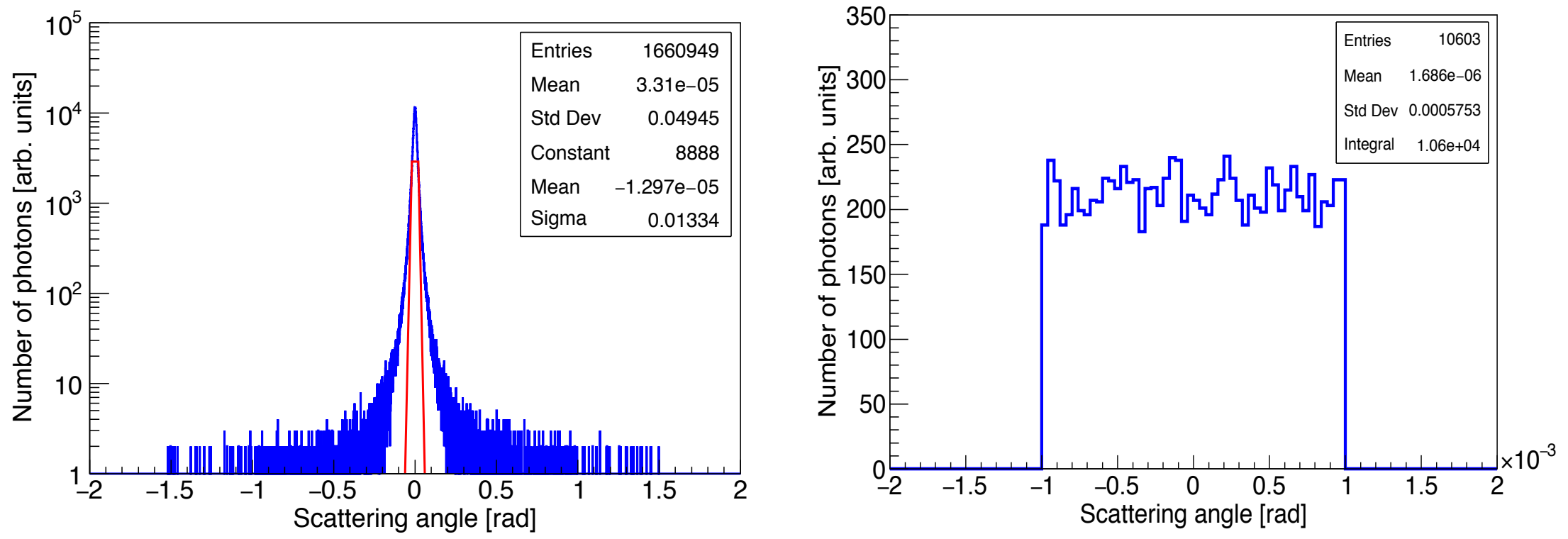

Figure 5-2: Angular distributions of photons emitted in the forward direction and in the detector with an acceptance of 2 mrad. Left: the angular distribution in the forward direction. Right: the angular distribution in the detector.

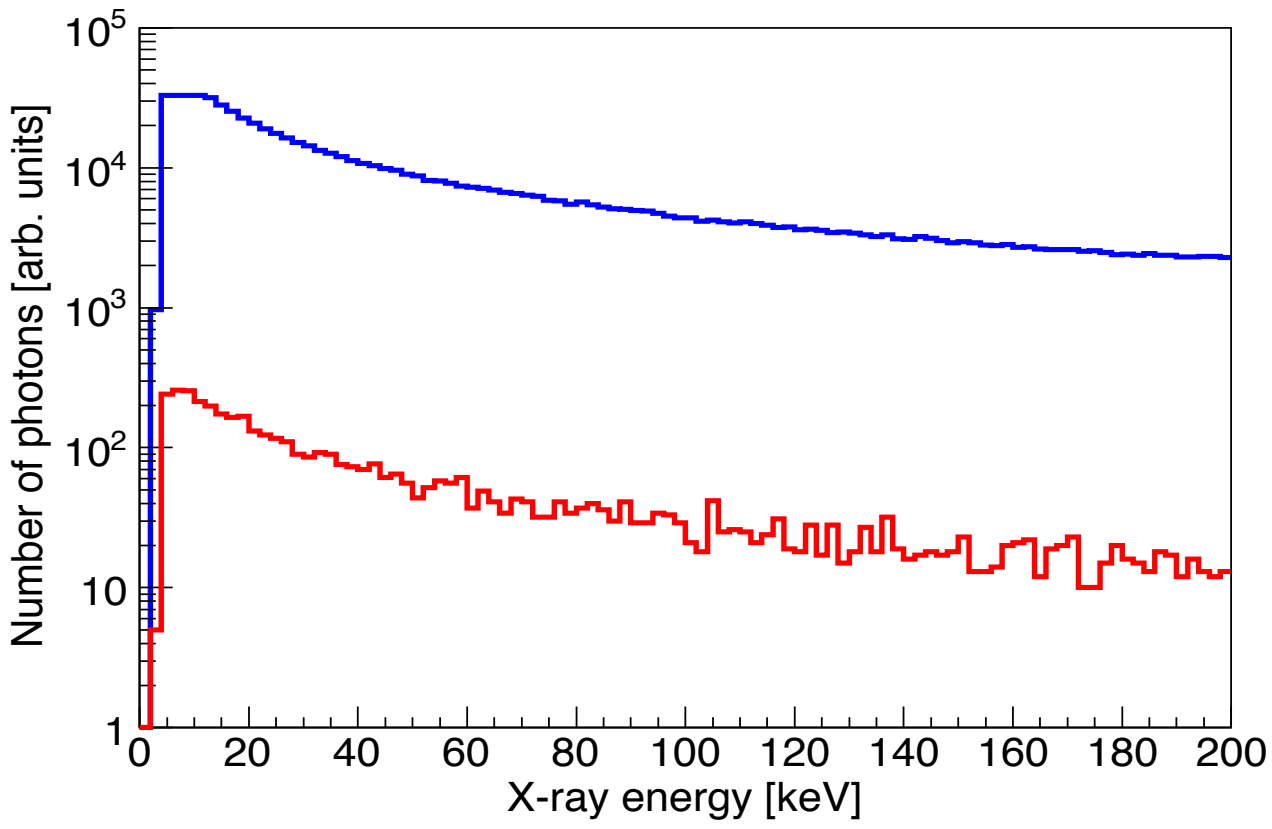

Figure 5-3: Bremsstrahlung spectra in the forward direction (blue) and in the detector (red).

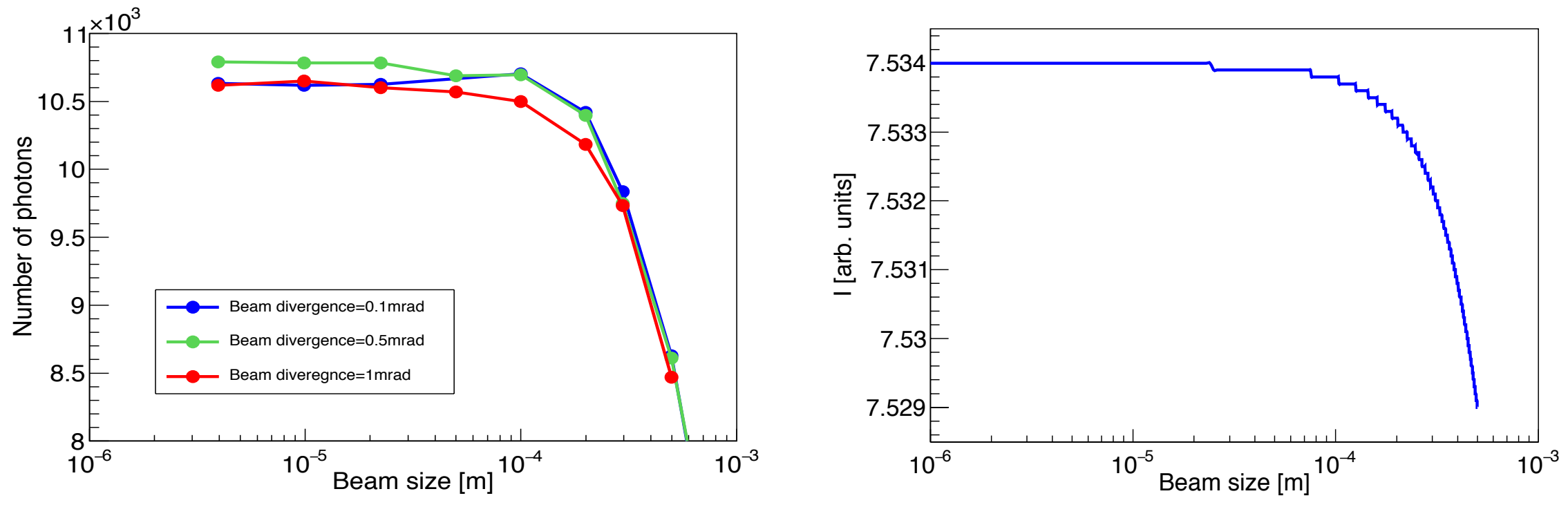

Figure 5-4: Number of photons hitting the detector depending on beam divergences and beam sizes of incident electrons. Left plot is simulation results with Geant4 and right plot is analytical calculation with Eq. (5-2).

Table 5-2: Number of photons generated in the forward direction for different electron energies of 43 MeV, 60 MeV, 80MeV, and 100 MeV.

| Electron energy MeV | Number of photons | Number of photons/Number of photons at 43 MeV |
|---|---|---|
| 43 | 1,660,949 | 1.00 |
| 60 | 1,711,277 | 1.03 |
| 80 | 1,745,297 | 1.05 |
| 100 | 1,764,527 | 1.06 |

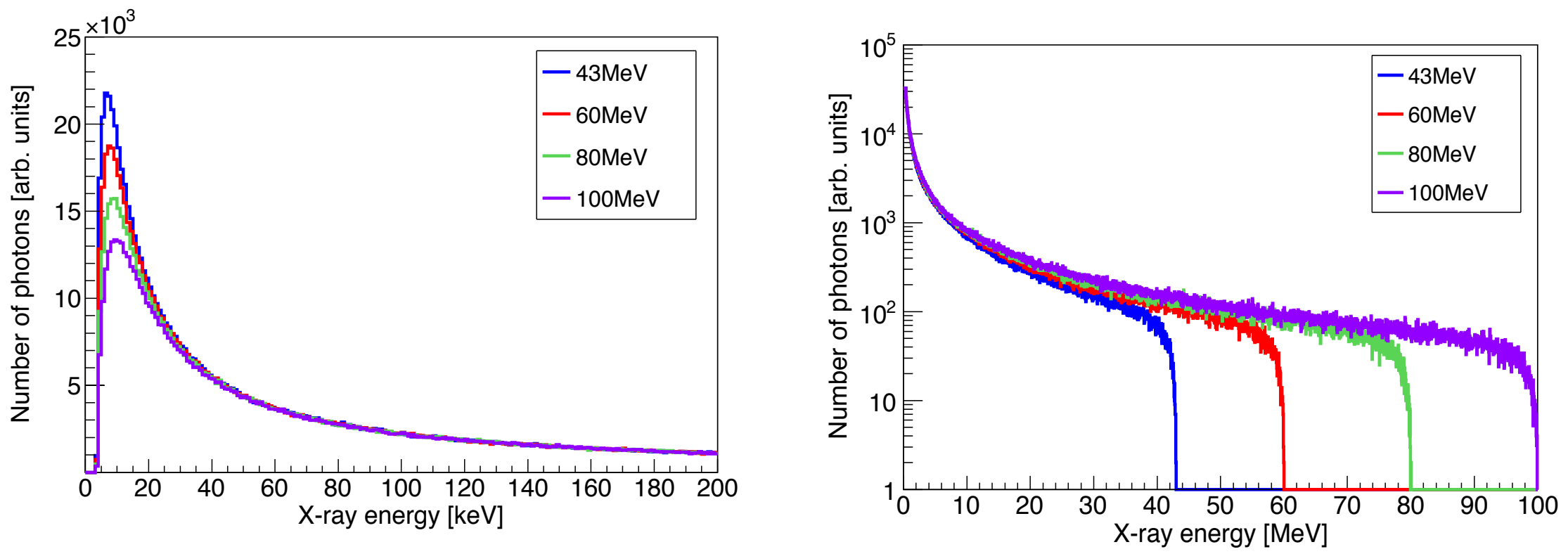

Figure 5-5: Bremsstrahlung spectra for different electron energies of 43 MeV, 60 MeV, 80MeV, and 100 MeV. Left: the energy spectra for 0-200 keV. Right: the energy spectra for 0-100MeV.

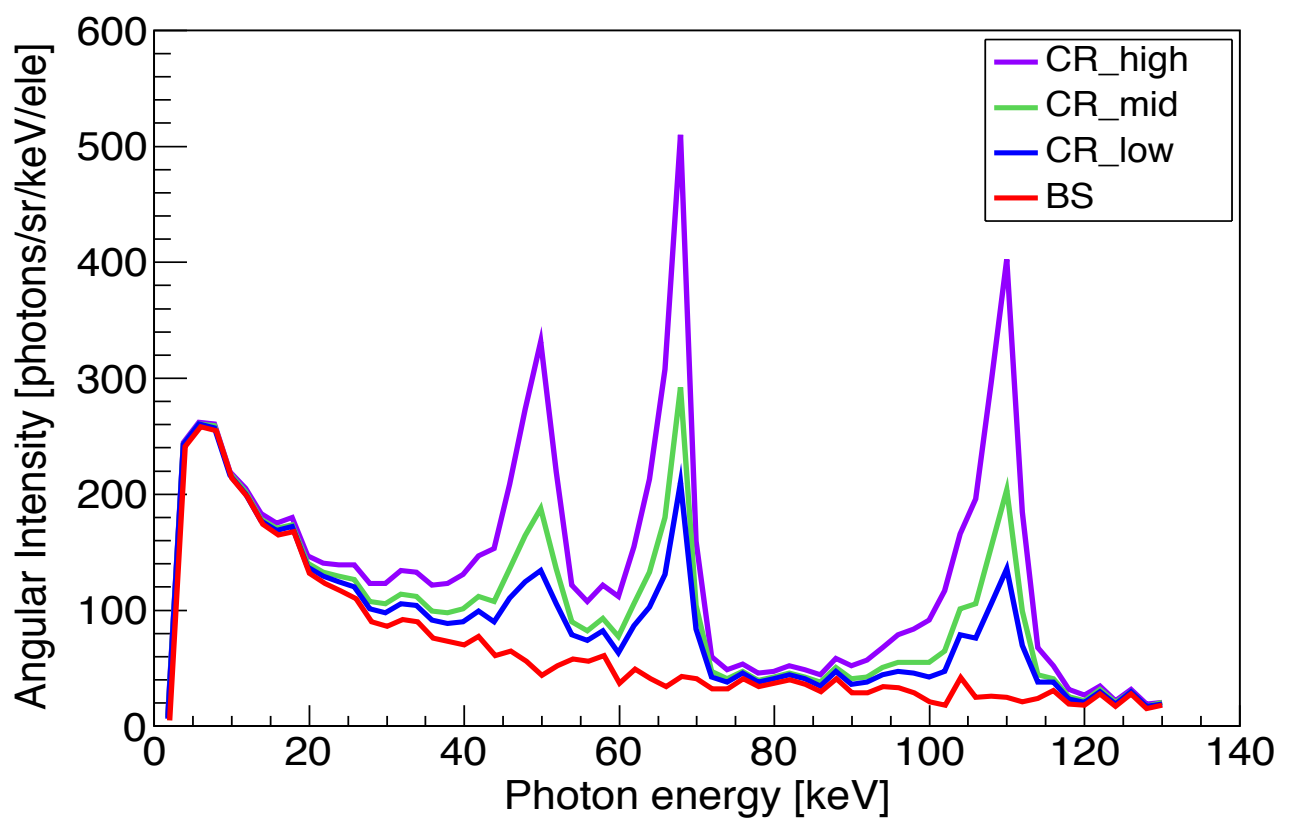

Figure 5-6: The expected X-ray spectrum including the background for difference estimates of the CR photon number.

## *6. Electron beam distributions after crystal*

This section describes electron beam sizes and beam divergences after the electron beam passes through the diamond crystal with and without channeling. This beam experiences multiple scattering events with atomic nuclei, and the beam divergence grows after passing through the crystal. Therefore, the electron beam emittance increases, which could cause particles loss from scraping at the beam pipe. The multiple scattering depends on whether the beam is channeled or not in the crystal. The rms scattering angle $\theta$ for an electron that is not channeled depends on the crystal thickness $L$, its atomic number, its mass number, and the electron momentum $p$, as [18,19]

$$\theta = \frac{13.6[\mathrm{MeV}]}{\beta c p}\sqrt{\frac{L}{L_{rad}}}\left[1 + 0.038\ln\left(\frac{L}{L_{rad}}\right)\right], \quad (6\text{-}1)$$

$$L_{rad} = \frac{716.4[\mathrm{g/cm^2}]\cdot A}{Z(Z+1)\ln(287/\sqrt{Z})} \quad (6\text{-}2)$$

where $\beta c$ is the velocity of the electrons, $L_{rad}$ the radiation length of the crystal, the mass number $A$, and the atomic number Z. The rms scattering angle after an electron of 43 MeV passes through a 168 μm thick diamond crystal can be computed to be about 8.3 mrad. However, the beam divergence for a channeled electron has been shown to be about 0.2 - 0.6 times smaller than that obtained from Eq. (6-1) [8]. Using recently added modules in Geant4 [16, 9], we estimated the scattering angles and energy spreads of the electron beam passing through the crystal under both channeling and non-channeling conditions.

The angular scattering of the electron after the crystal calculated using Geant4 is shown in Figure 6-1. The initial beam divergence at the crystal is set to be 0.1 mrad, and $10^5$ electrons are used in both cases. The left plot in Fig. 6-1 shows the angular distribution of electrons at the crystal entrance and the right plot shows the distribution after the crystal with and without channeling. The standard deviation of the entire distribution (including the tails) without channeling is about 10.1 mrad, about 100 times larger than that before the crystal. On the other hand, the standard deviation of the distribution after channeling is 7.7 mrad, or approximately 0.8 times the value without channeling. The beam divergence after the electron beam is channeled in the crystal become smaller than that after without channeling. Figure 6-2 shows the beam sizes of electrons at a monitor in front of the detector, after channeling and without channeling in the crystal. The beam monitor is assumed to be at 1.5 m from the crystal. Without channeling in the crystal, the beam sizes at the monitor are (X, Y) = (15.4 mm, 15.2 mm) while with channeling, they decrease to (11.3 mm, 11.1 mm). There is a small shift in the horizontal position of the centroid with and without channeling, but this could be an artifact of the relatively small number of particles used in the simulation.

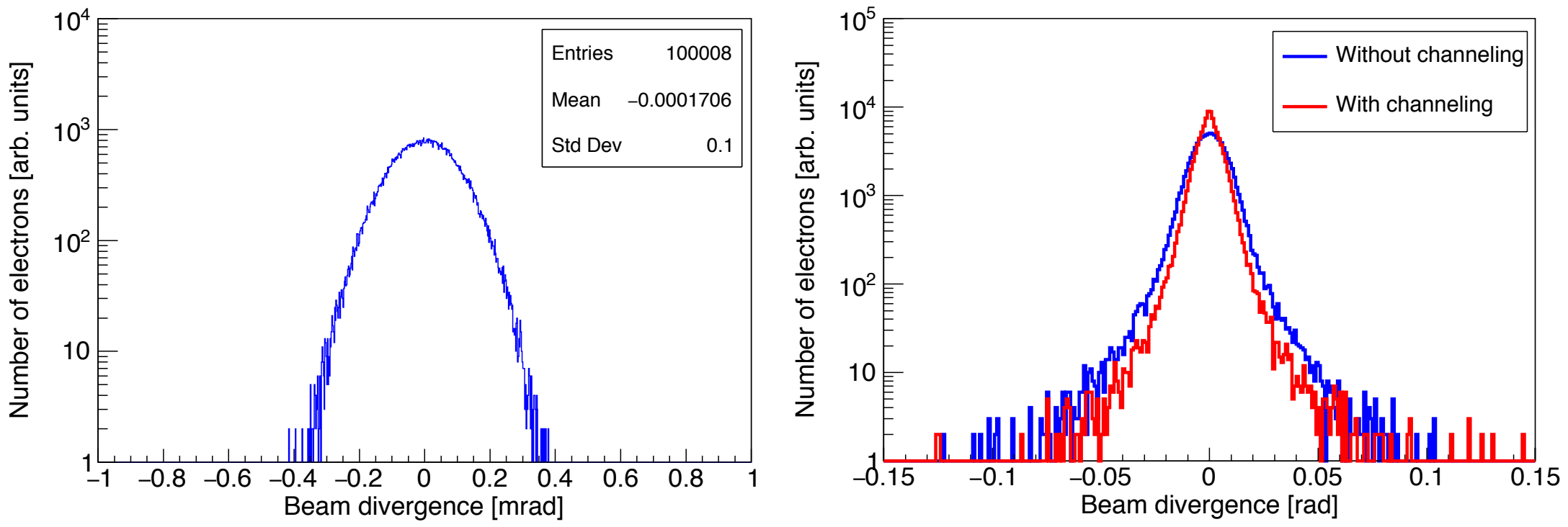

Figure 6-1: Angular distributions of a 43 MeV electron beam before and after going through the diamond crystal with the thickness of 168 μm. Left plot [mrad scale]: the distribution at the crystal entrance. Right plot [rad scale]: the beam distribution with and without channeling. Standard deviations with and without channeling are 7.7 mrad and 10.1 mrad, respectively.

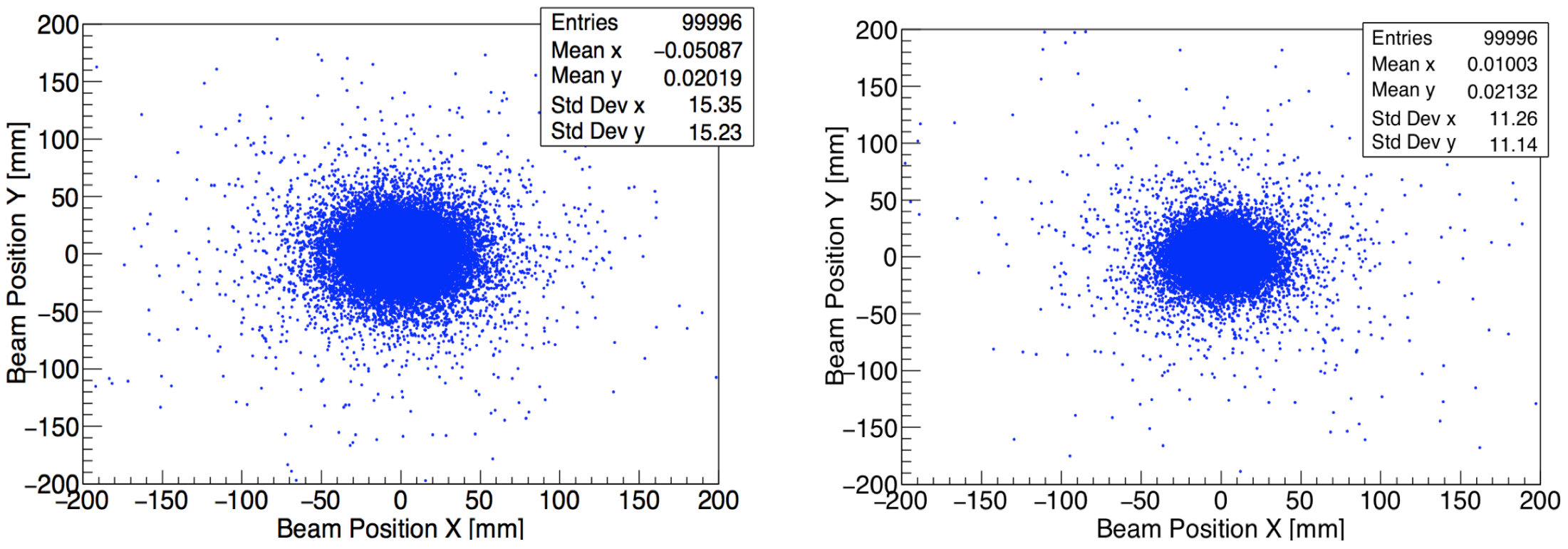

Figure 6-2: Beam positions and sizes (x, y) after the non-channeled (left) and the channeled (right) electron beam in the diamond crystal.

When relativistic electrons pass through the crystal, the energy loss is caused by the ionization loss and radiation loss (bremsstrahlung). The critical energy $E_c$ when the two energy losses are equal is written as [16]

$$E_c[\mathrm{MeV}] = \frac{610\ \mathrm{MeV}}{Z + 1.24}, \qquad (6-3)$$

where $Z$ is the atomic number of the crystal. This equation means that the ionization loss is dominant for an electron beam with energy below the critical energy. The critical energy $E_c$ for a diamond crystal of Z=6 is 84 MeV, which means that the ionization loss contributes strongly to the electron energy loss in our CR experiment with a 43 MeV electron beam. The ionization loss can be analytically calculated by the Bethe-Bloch equation [16]

$$-\frac{dE}{dx} = 0.1535[\mathrm{MeV/gcm^2}]\frac{Z}{A}\frac{\rho}{\beta^2}\left[\frac{1}{2}\ln\left(\frac{2m_ec^2\beta^2\gamma^2T_{max}}{I^2}\right) - \beta^2 - \frac{\delta(\beta\gamma)}{2}\right], \qquad (6-4)$$

$$T_{max} = \frac{2m_ec^2\beta^2\gamma^2}{1 + 2\gamma m_e/M + (m_e/M)^2}$$

$$I \approx 16Z^{0.9}[eV]$$

where $\rho$ the crystal density, $\beta c$ the velocity of a electron beam, $Z$ atomic number, $A$ atomic mass, $m_e$ electron rest mass, $I$ the excitation potential, and $\delta$ density effect correction. An experimental energy loss for a thin crystal would be smaller than that calculated by Eq. (4-4), because the Bethe-Bloch equation describes the mean energy loss and the energy loss obtained from experiments follows the Landau distribution [18]. Using Eq. (6-4), the mean energy loss for an electron beam of 43 MeV and a 168 μm thick diamond crystal can be computed to be about 100 keV, about 0.2 % of the initial electron energy. Figure 6-3 shows the electron energy distributions after channeling and without channeling calculated with Geant4. The green curve in Fig. 6-3 is the initial energy distribution of electrons with an energy spread of 0.1 %, the blue curve is the energy distribution after the crystal without channeling, and the red curve is after channeling. For the case without channeling, the energy distribution has a peak at 42.9 MeV and the energy loss is about 100 keV. On the other hand, for the case with channeling, the peak is at 42.95 MeV and the energy loss is about 50 keV.

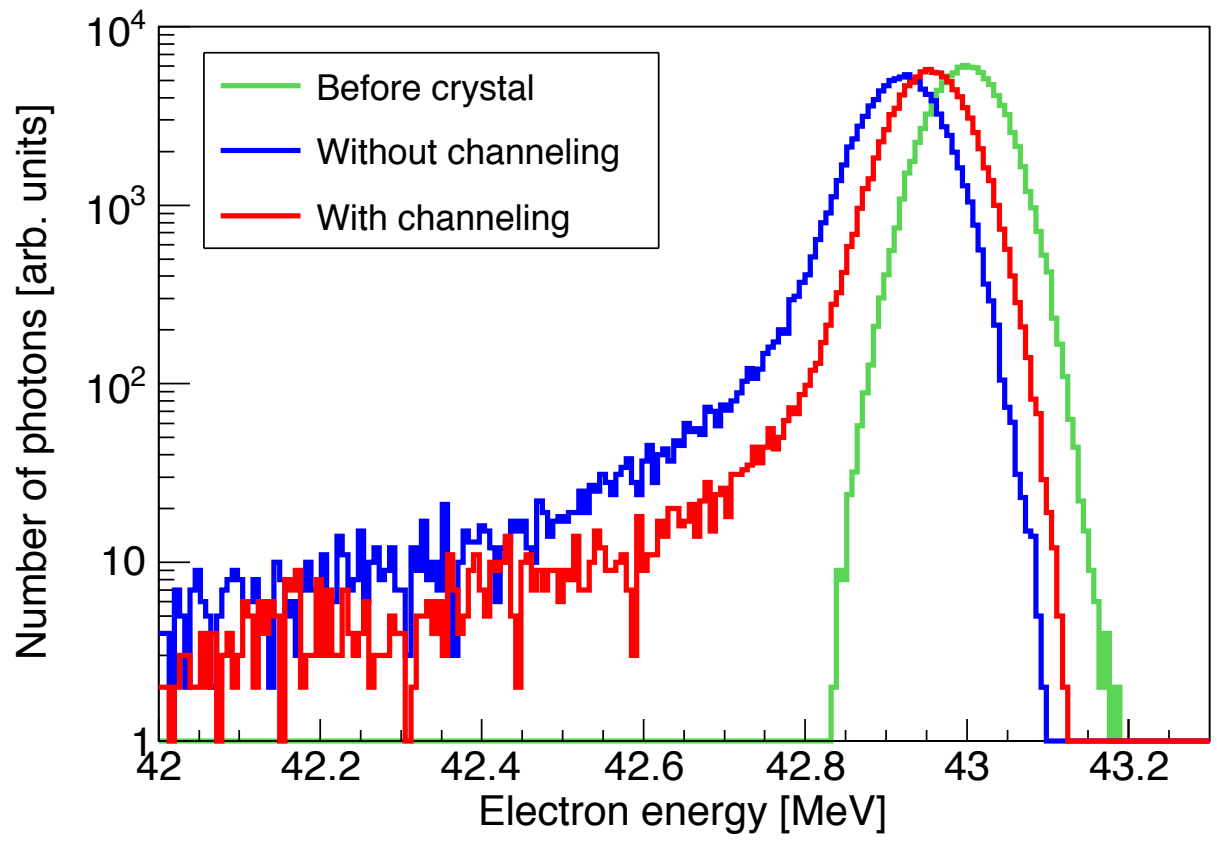

Figure 6-3: Energy distributions of the electron beam before and after the diamond crystal. The green curve shows the initial energy distribution. The red and blue curves are the distribution with channeling and without channeling, respectively.

## *7. Compton scattering for X-ray detector*

The experimental layout in Fig 3-1 shows an X-ray detector downstream of the crystal. This configuration will be used at low bunch charges when the photon rate emitted in to the detector's acceptance is about 1 photon per bunch. At higher bunch charge and the nominal laser frequency of 3 MHz, the photon rate will be high enough to cause photon pileup in this detector. Two or more photons arriving within the detector's response time (about 25 μs) will be registered as a single photon with an energy which is the sum of all the photon energies leading to a wrong spectrum. In this section, we discuss the use of a second X-ray detector placed orthogonal to the beamline which will detect photons Compton scattered from a plastic plate in order to avoid this pile up effect.

In our experiment, with a laser pulse repetition rate of 3 MHz, and for 1 pC/bunch, the number of electrons going into the diamond crystal per second can be computed using Table 5-1

$$600\ [photons] \times 3\times10^{6}[\mathrm{Hz}] = 1.8\times10^{9}\ [\mathrm{photons/s}]. \quad (7-1)$$

Table 7-1 shows the expected photon counts per second hitting the detector for each charge. This estimate considers only the BS background, which indicates that the pile up will be caused even when a bunch charge of 1 pC is used.

For the pile up rejection, it is important to reduce the photon flux going into a detector. Commonly, for low photon energies, the number of photons hitting a detector can be reduced using an attenuator such as Al and Brass, and we can extract the true photon rate by calculating the number of photons absorbed into the attenuator with the absorption data which is supplied by NIST [21]. By contrast, for high photon energies, using attenuators is not effective due to a low absorption cross sections at high energies. This will be true in our experiment with expected CR photon energies ranging from 50 keV to 110 keV. Thus, we will utilize Compton scattering that can significantly reduce the photon rate for high photon energies.

Compton scattering results from the interaction of a photon with free electrons in a material substance. The scattered photons experience energy loss, resulting in shifts to longer wavelength. The differential cross section for the Compton scattering is given by the Klein-Nishina formula:

$$\frac{d\sigma}{d\Omega} = \frac{r_e^2}{2}\frac{1}{\left[1 + h\nu_0[1-\cos\theta]\right]^2}\left(1+\cos^2\theta + \frac{(h\nu_0)^2(1-\cos\theta)^2}{1+h\nu_0(1-\cos\theta)}\right), \quad (7-3)$$

$$h\nu = \frac{h\nu_0}{1+\left(\frac{h\nu_0}{m_e c^2}\right)(1-\cos\theta)}, (7-4)$$

where $r_e$ is the classical radius of the electron, $h\nu_0$, $h\nu$ are the energies of the incident and scattered photons respectively and $\theta$ is the scattering angle. In order to know the relation of a photon's cross section, scattering angle, and final energy, the differential cross section and photon energy as a function of scattering angle for incident photon energies of 50 keV, 70keV, and 110 keV are plotted in Figure 7-1 using Eq. (7-3) and (7-4). Most photons are scattered in the forward and backward direction to the incident photons. Since the differential cross section has a minimum at 90 degrees, the detector should be placed orthogonal to the beamline (incident photons). Figure 7-2 shows the layout of the second X-ray detector. X-ray energies of 110 keV, 70 keV, and 50 keV scattered at 90 degrees have their energies shifted to 90 keV, 60 keV and 45 keV, respectively, because the photons lose energy to the scattering electrons.

Table 7-1: The number of background photons per second registered in the detector for three charges.

| Charge [pC] | Photons/s |
|---|---|
| 1 | $1.8 \times 10^{9}$ |
| 20 | $3 \times 10^{10}$ |
| 200 | $3 \times 10^{11}$ |

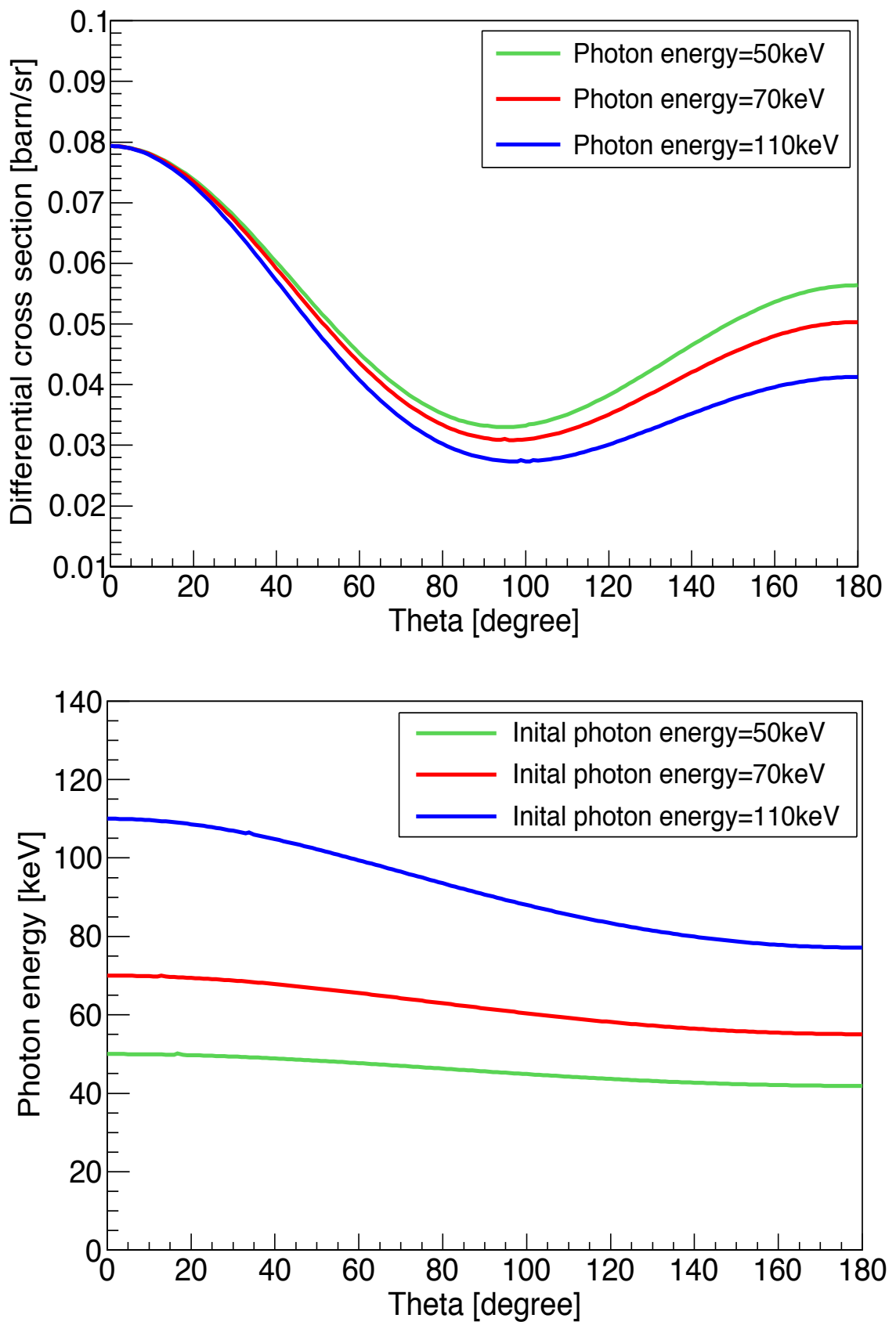

Figure 7-1: Differential cross-section (upper) and photon energy (lower) depending on scattering angle.

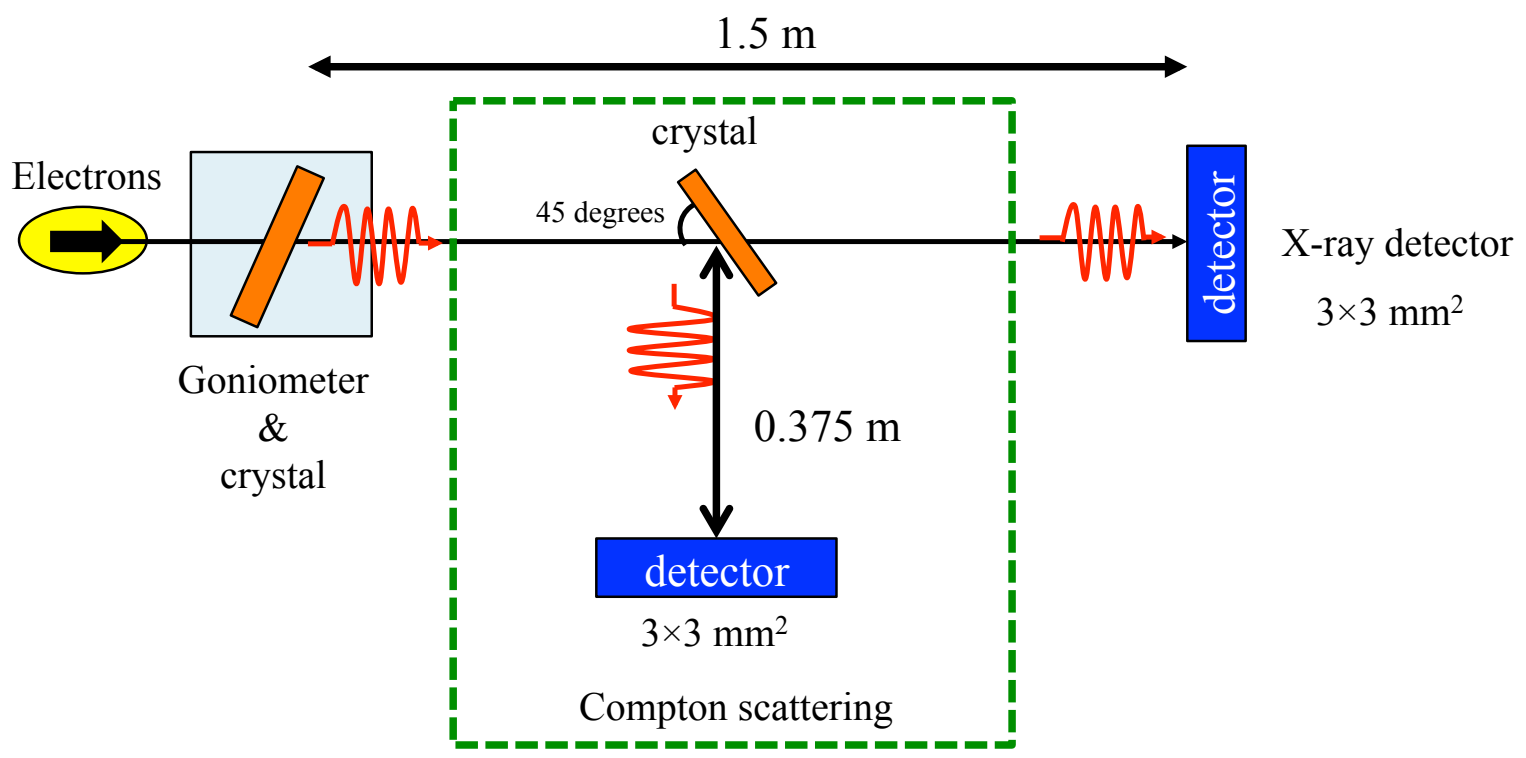

Figure 7-2: Layout of the detector.

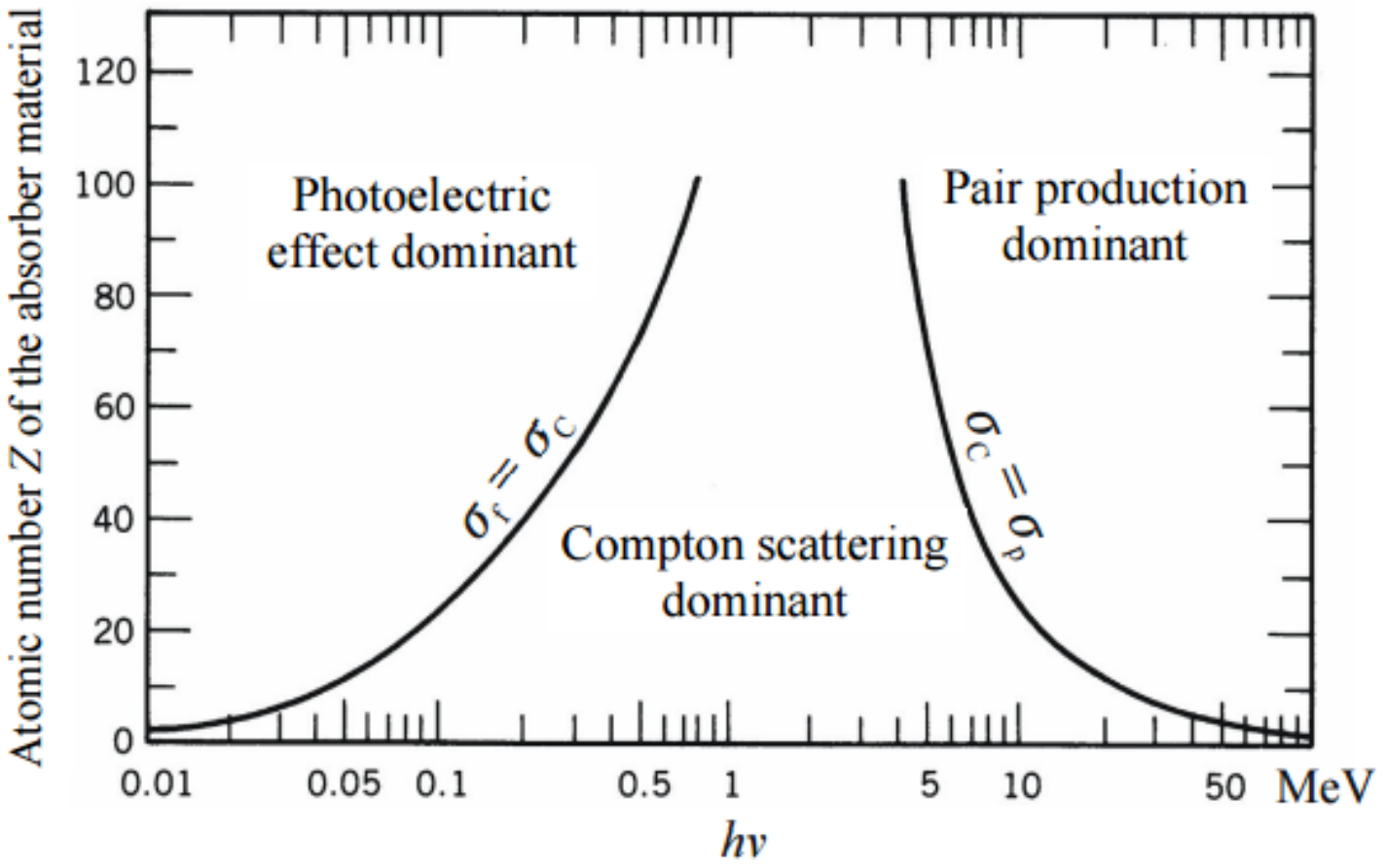

Figure 7-3: Three major types of photon interaction with matter [22]. The curves demarcate the regions where each effect is dominant.

Figure 7-3 shows the three major types of photon interactions with matter [22]. Since photons with energies of 110, 70, and 50 keV are expected in our CR experiment, the target material should have an atomic number below 30 owing to the dominance of Compton scattering in this energy range for these materials. In order to select appropriate materials to use in our experiment, we have tried diamond, Al, Si, PMMA, and PVC plates, each with a thickness of 2 mm. Photon counts in the X-ray detector were simulated using Geant4. As an initial condition, $10^9$ photons with an energy of 70 keV, and an energy spread of 10% was used. The detector with the aperture of 3x3 $mm^2$ is assumed to be located at 0.375 m from the Compton scattering plate, see Fig 7-2. Figure 7-4 shows the number of Compton scattered photons and their energies hitting the detector for the different materials. Photons with energies of 60 keV and energy spreads of 10% are scattered in the detector; the simulated energies agree with the analytical calculation with Eq. (7-4). The Compton scattered photon count decreases with an increase of the atomic number Z due to the absorption within the material. For the organic materials, the number of photons in the detector decreases by seven orders of magnitude, therefore, we decided to use PVC or PMMA as the material for Compton scattering.

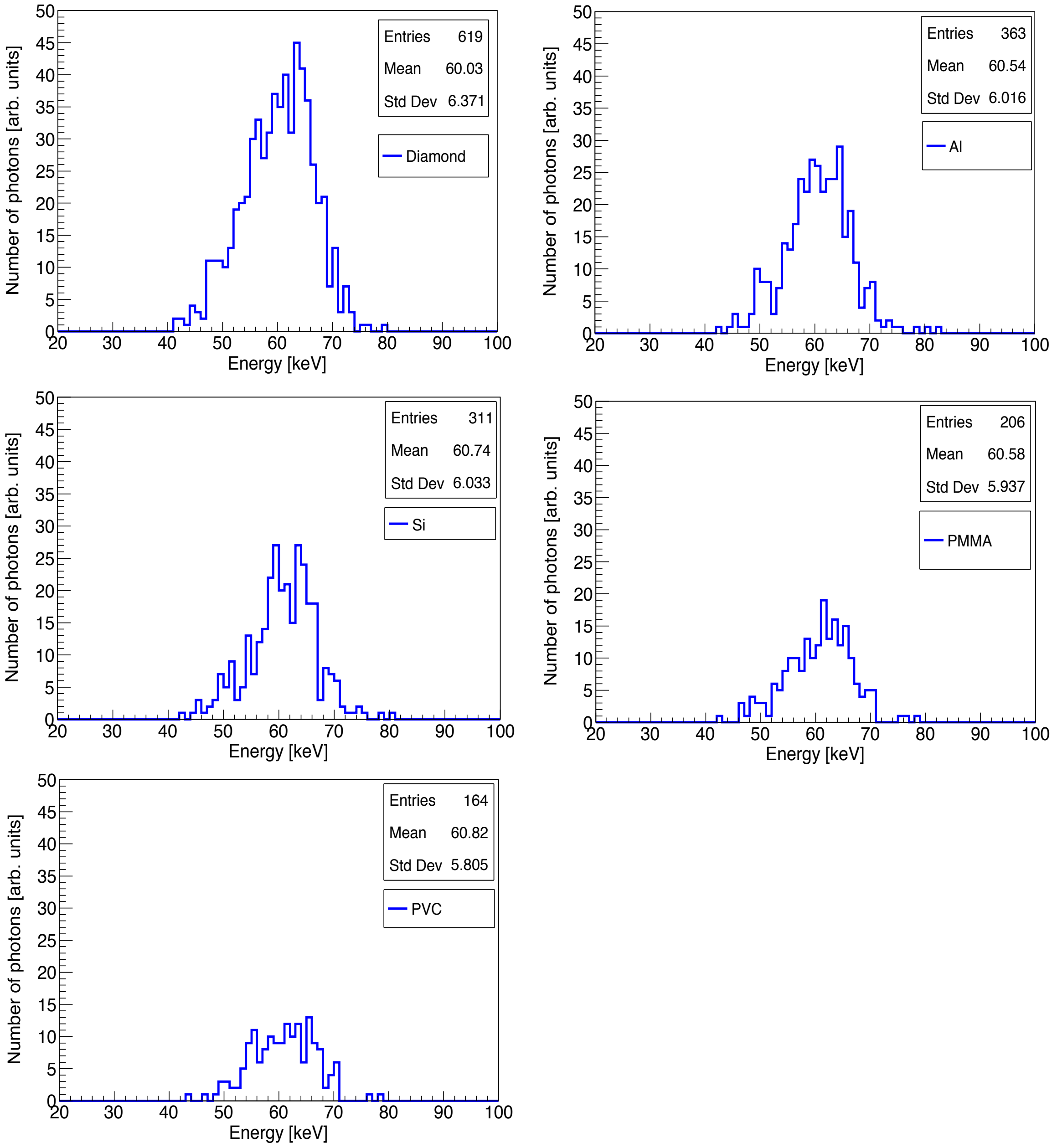

Figure 7-4 The number of Compton scattered photons and the energy in the detector for different crystals.

## *8. CONCLUSIONS*

In this paper, we have studied various aspects related to channeling radiation (CR) experiments planned at Fermilab's FAST photoinjector. The topics discussed include: 1) beam optics for both a high brightness and high yield solutions, 2) CR photon yields including background from bremsstrahlung (BS), 3) electron beam distributions with and without channeling through a diamond crystal, and 4) an X-ray detector system to avoid pile-up.

The beam optics for the two operations requires different solutions: low beam sizes for high brightness X-rays and low beam divergences for high yield X-rays. We simulated optics solutions for three bunch charges of 1 pC, 20 pC, and 200 pC with the SAD code. Small beam sizes require strong focusing and the impact of chromatic aberrations become more significant as the bunch charge increases due to a corresponding increase in the momentum spread. Thus for the high brightness solution, the minimum rms beam sizes at the diamond crystal were (1, 4, 10) μm respectively for the three bunch charges studied, As for the high yield solution, we obtained beam optics solutions with a beam divergence of 0.1 mrad at the crystal for the different bunch charges. This divergence is much less than the critical angle for channeling, about 1 mrad at 43 MeV.

In order to examine the effect of the BS background on the CR signal, we simulated the BS generated from electrons propagating in the diamond crystal using Geant4. The signal to background ratio varies from 2 at the 50 keV CR spectrum line to 7 at the 110 keV CR line.

Electron beam distributions after propagating through the crystal with and without channeling were simulated with Geant4. The beam divergence after channeling was 0.8 times smaller than the case without channeling. Also, under channeling conditions, the electron beam size after the crystal was smaller in both planes. These results suggest that a smaller emittance growth and beam size after the crystal could be used to test for channeling in the initial stage of the CR experiments.

To avoid pile up and saturation of the X-ray detectors, we utilized a second X-ray detector orthogonal to the beam line and Compton scattering to reduce the photon count. Geant4 simulations show that a 2 mm thick scattering plate made of organic matter such as PVC or PMMA can reduce the photon number in this detector by seven orders of magnitude.

## *Acknowledgements*

We are grateful to Dr. E. Bagli (INFN) for supplying Geant4 applications for electron channeling. We thank Dr. D. Mihalcea (NIU) for helpful discussions and suggestions. The first author would like to express his gratitude to SOKENDAI (The Graduate University of Advanced Studies) and KEK for The Short-Stay Study Abroad Program.